\magnification=1200

\pageno=1

\centerline {\bf THE GENERALIZED  MOYAL-NAHM  AND  }

\centerline {\bf CONTINUOUS  MOYAL TODA  EQUATIONS   }

\medskip

\centerline { Carlos Castro$^*$ , Jerzy Plebanski$^+$ }

\centerline { $^{*+}$ Dept. de Fisica, Centro de Investigacion y Estudios Avanzados del IPN  }

\centerline { Apdo. 14-740 CP 07000 , Mexico DF; Mexico }

\centerline { $ ^*$ Center for Theoretical Studies of Physical Systems}

\centerline { Clark Atlanta University, Atlanta,  Georgia 30314 USA}

\smallskip

\centerline { September  97. Revised April. 98 } 

\smallskip

\centerline {\bf ABSTRACT}

\smallskip

We present in detail  a class of solutions to the  $4D~SU(\infty)$ Moyal  Anti 
Self Dual Yang Mills
equations that are related to  $reductions$ of the generalized
Moyal Nahm quations using the Ivanova-Popov ansatz. The former 
 yields solutions to the ASDYM/SDYM equations for arbitary gauge groups.  
A further dimensional  reduction
yields solutions to the Moyal   Anti Self Dual Gravitational 
equations. The Self Dual Yang Mills /Self Dual Gravity  case
requires a separate study. The $SU(2)$ and $SU(\infty)$ ( continuous)  Moyal Toda
equations are derived and solutions to the latter equations  in 
$implicit$  form are proposed
via the Lax-Brockett double commutator formalism . An explicit map taking the Moyal heavenly form ( after a rotational Killing symmetry reduction)  into the $SU(2)$ Moyal Toda  field is found. Finally,  the generalized
Moyal Nahm equations are conjectured  that 
contain the continuous $SU(\infty)$ Moyal Toda equation after a
 suitable reduction.  Three different embeddings of the three different types of Moyal Toda equations
into the Moyal Nahm equations  are
 proposed.

\smallskip


\smallskip

\centerline {\bf I. Introduction}

The quantization program of the $3D$ continuous Toda theory ( $2D$ Toda molecule
) is a challenging enterprise that we believe would enable us  to understand 
many of the features of the quantum dynamics and spectra of the quantum self 
dual membrane [1]. 
This is based on the observation  that the light-cone-gauge ( spherical)  
supermembrane moving in a $D$ dimensional  flat spacetime background has a 
correspondence with a $D-1$ $SU(\infty)$ Super Yang-Mills theory , 
dimensionally reduced to one temporal dimension; i.e. with a $SU(\infty)$ 
supersymmetric gauge quantum mechanical model ( matrix model) [2,6] .

It was shown in [1] that exact ( particular) solutions of  the $D=11$ light-cone
( spherical) supermembrane  
, related to the $D=10$ $SU(\infty)$ SYM theory, could  be constructed based on 
a particular class of reductions of the SYM equations from higher dimensions to 
four dimensions [3].  In particular, solutions of the $D=10$ YM equations given 
by the $D=10$ YM potentials , ${\cal A}_\mu$,  can be obtained in terms of the 
$4D$ YM potentials, $A_i$,  that obey the $D=4$ Self Dual YM equations.
Dimensional reductions of the latter $SU(\infty)$ SDYM equations  to one 
temporal dimension are equivalent to the $SU(\infty)$ Nahm equations in the 
temporal gauge $A_0=0$.

Finally, the embedding of the continuous $SU(\infty)$ Toda equation into the 
$SU(\infty)$ Nahm equations was performed in [1] based on the 
connection between the $D=5$ self dual membrane and the $SU(2)$ Toda 
molecule/chain equations  [4]. A continuous Toda theory in connection to self dual gravity  was al
so found by Chapline and Yamagishi [5]  in the description of  a 
three-dimensional  version of anyon superconductivity. 
Based on the theory of gravitational instantons a $3D$ model describing the 
condensation of quasiparticles ( $chirons$ ) 
with properties related to fractional statistics was found.

The classical Toda theory can be 
obtained also from a rotational Killing symmetry reduction [15] of the $4D$ Self 
Dual 
Gravitational (SDG)  equations expressed in terms of first heavenly 
form that furnish ( complexified) self dual metrics of the form :
$ds^2 =(\partial_{x^i}\partial_{{\tilde x}^j}\Omega) dx^i d{\tilde x}^j$ for 
$x^i=y,z$; ${\tilde x}^j ={\tilde y},{\tilde z}$ and $\Omega$ is the 
first heavenly form. The latter equations can, in turn, be obtained from a 
dimensional reduction of the $4D~SU(\infty)$ Self Dual Yang Mills equations 
(SDYM), an effective $6D$ theory [7,8] and references therein. 
The Lie algebra $su(\infty)$ was shown to be isomorphic 
( in a basis dependent limit) to the Lie algebra of area preserving 
diffeomorphisms of a $2D$ surface, $sdiff~(\Sigma)$ by Hoppe [2]. It is for this
reason that a WWM quantization of the reductions of the first heavenly equation will be  used in
this letter.

Using our results of [9] based on [10] we have shown that a Weyl Wigner Groenwold Moyal ( WWGM)   [11] 
quantization approach yields a straighforward quantization scheme for the 
$3D$ continuous Toda theory ( $2D$ Toda molecule). Supersymmetric extensions 
can be carried out following [7] where we wrote down the supersymmetric analog 
of the heavenly  equations for SD Supergravity.

There are some  differences between our results and those which in general
have appeared in the literature . Among these are  (i) One is $not$ taking the 
limit of $\hbar \rightarrow 0$ while having $N=\infty$ in the classical 
$SU(N)$. Recently, Fairlie [23] has written solutions to Moyal-Nahm equations, with  $\hbar \not=0$ 
for the eight tranverse membrane coordinates in $D=11$ in terms of spinors using the WWM formulation.   
(ii) We are working with the generalized Moyal-Nahm equations involving a Moyal
bracket w.r.t an $enlarged$ phase space , $q,p,q',p'$ and not with the 
standard $SU(2)$ Moyal-Nahm equations involving a Moyal bracket w.r.t 
$q,p$ only.  We have $\hbar \not= 0;N=\infty$ 
simultaneously. (iii) The connection with the self dual membrane and 
$W_{\infty}$ algebras was proposed in [1].    
The results of [10] become very useful in the implementation of the 
WWM quantization program and in the embedding of the $SU(2)$ 
Moyal-Nahm solutions into the generalized  Moyal-Nahm equations studied 
in the present work.

In the next section we  present in detail  a class of solutions to the  $4D~SU(\infty)$ Moyal  Anti 
Self Dual Yang Mills
equations that are related to  $reductions$ of the
generalized Moyal Nahm quations using the Ivanova-Popov ansatz. The former 
yields solutions to the ASDYM/SDYM equations for arbitary gauge groups.  
A further dimensional  reduction
yields solutions to the Moyal   Anti Self Dual Gravitational 
equations. The Self Dual Yang Mills /Self Dual Gravity  case
requires a separate study.

In {\bf III} the  $SU(2)$ and $SU(\infty)$ Moyal Toda ( continuous) 
equations are derived and solutions to the latter equations  in 
$implicit$  form are proposed
via the Lax-Brockett double commutator formalism of the continuous Toda
equation [13] . By $SU(\infty)$ Moyal  one means the Moyal deformations of the symplectic diffeomorphisms in $4D$ instead of $2D$. This is  explained in detail in the text. The master Legendre transform between solutions to the rotational Killing
symmetry reductions of the Moyal heavenly equations and the Moyal Toda equations is studied. In particular, the explicit map taking the Moyal heavenly form into the $SU(2)$ Moyal Toda field is found. Finally, in
{\bf IV},  the  generalized Moyal Nahm equations are provided that 
$contain$ the continuous $SU(\infty)$ Moyal Toda equation after a
suitable reduction.  Three different  embeddings  of the different types of Moyal Toda equations into
the Moyal Nahm equations  are provided. At the end of {\bf IV} we show explicitly how to connect the Moyal heavenly form with the continous Toda field via the Moyal-Lax pair formalism. The remaining obstacle is to write down the 
explicit Moyal quantization of all the equations involved in the Ivanova-Popov construction [3]. This is necessary in order to write down the equation governing the deformations of the scalar field involved in the construction of [3] , and which is mapped into the Moyal deformed continuous Toda field via deformations of the twistor transform. Deformations of twistor surfaces have yet to be constructed. For comments in that direction we refer to the work of Strachan and Takasaki [16,29] in their  study of higher dimensional integrable models.

The most salient
feature of the generalized  Moyal-Nahm equation  is that it involves 
an effective $8D$ theory associated with symplectic diffeomorphisms of a $4D$
manifold. This $8D$ effective theory may have an important role in understanding
the quantum dynamics of the $11D$ supermembrane based on the  the 
M(atrix) Membrane models and their integrability properties [6,23].

In the conclusion we discuss ( among other things) briefly how deformation quantization techniques
for higher dimensional volume forms, the Zariski quantization [22], may be used
to quantize $p$ branes. Other types of deformations are possible, like  
those which give up the $associative$ character of the Moyal product, the
so-called $q$ deformations, and which can be used to construct  deformations  of the self 
dual membrane we refer to [24]. 
We expect that a $q$-Moyal deformation program of the self dual membrane 
might 
yield important information about how to quantize  the full  
membrane theory beyond the self dual exactly integrable sector and that the 
particle/soliton spectrum of the underlying  Conformal Affine Toda models will shed some 
light into the particle content of the more general theory [1,9]

Note : 

Sometime after this work was completed we were informed [27] based on Strachan's work [16] that a map from the master equation involving the Moyal deformations of the Self Dual Yang-Mills equations  to Strachan's $SU(2)$ Moyal Toda equation ( which in reality is a Toda lattice model whose discrete spacing is given by multiples of $\hbar$)  could be obtained by a suitable dimensional reduction. However, the latter is $not$ a rotational Killing symmetry 
reduction; i.e  the particular dimensional reduction of the effective $6D$ master equation is $not$ a Moyal deformation of the heavenly equation ( although the heavenly equation can be obtained from the master equation) . The results of [27] are based on a $two$  step process where $after$  performing  the WWGM map and explicit introduction of $\hbar$ is put in by hand. More on this shall be explained in the text.

\smallskip

\centerline {\bf II. The $SU(\infty)$ MOYAL ASDYM/SDYM equations}

\smallskip

We will study in this section the Moyal $4D~SU(\infty)$ ASDYM/SDYM 
in connection to the ASDG/SDG equations and the continuous $2D$
Toda molecule. To begin with, the Moyal bracket of two YM potentials $A_y, A_{{\tilde y}}$, for
example, can be expanded in powers of $\hbar$  as  [16] :

$$ \{A_y, A_{{\tilde y}}\}_{q,p}\equiv \sum_{s=0}^\infty {(-1)^2\hbar^{2s}\over
(2s+1)!}  \sum_{l=0}^{2s+1} (-1)^l (C^{2s+1}_l )[\partial_q^{2s+1-l}\partial_p^l
A_y] [\partial_p^{2s+1-l}\partial_q^l A_{\tilde y}]. \eqno (1)$$
where $C^{2s+1}_l$ are the binomial coefficients. 

The crucial difference between the 
solutions of the $SU(2)$ Moyal-Nahm eqs [10] and the generalized  Moyal-Nahm 
case is that one $must$ have an $extra$ explicit dependence on another $set$ of 
phase space variables, $q',p'$. 
In particular, those $reductions$ of the generalized  Moyal Nahm equations
that are linked to the Moyal Toda equations must have an extra $t$ dependence
for the YM potentials.  
The continuous Toda molecule equation as well as the usual Toda system  may be
written in the double commutator form of the Brockett equation [13] :

$${\partial L (\tau,t) \over \partial \tau} =[L,[L,H]]. \eqno (2)$$
$L$ has the form 

$$L\equiv A_{+} + A_{-} =X_o (-iu)+X_{+1} (e^{(\rho/2)})+X_{-1} (e^{(\rho/2)}). \eqno (3)$$
with the connections $A_{\pm}$ taking values in the subspaces 
${\cal G}_o \oplus {\cal G}_{\pm 1}$ of some {\bf Z}-graded continuum Lie algebra ${\cal G}=\oplus _m
{\cal G}_m $ of a novel class.  
$H=X_o (\kappa)$ is a continuous limit of the Cartan element of the principal 
$sl(2)$ subalgebra of ${\cal G}$. The functions $\kappa (\tau,t),u(\tau,t),\rho(\tau,t)$ 
satisfied certain equations given in [13].

Upon the elimination of $u$ one obtains the Toda equation. A naive look at (2) 
might beg the question : Where is the $t$ dependence ? The $t$ dependence is 
$encoded$ in the continuum algebra commutation relations of the generators that
are parametrized by a family of functions depending on both $\tau,t$. It is in
this fashion that the $t$ dependence makes its presence in (2).

To implement the Weyl Wigner Moyal (WWM) prescription , one may consider the case when   {\bf G}  
is a group of unitary operators acting in the Hilbert space of square integrable
functions on the line, 
$L^2 (R^1)$. Then, ${\cal G}$ is now the associated ( continuum ) Lie algebra of
self-adjoint   
operators acting in the Hilbert space, $L^2 (R^1)$. 
The following  operator-valued  quantities depend  on the two
coordinates, $\tau,t$  and obey  
the operator version of the 
Brockett equation  and after the WWM  map reads :

 $${\partial {\hat L} (\tau,t) \over \partial \tau} =
{1\over i\hbar}  [ {\hat L},  {1\over i\hbar} [ {\hat  L},{\hat H}]]. 
\leftrightarrow  {\partial { \cal L}  \over \partial \tau} =
 \{ { \cal   L},  \{ { \cal  L},{ \cal H}\}\}. \eqno (4)$$
where ${\cal L} (\tau,t,q,p;\hbar),{\cal H}(\tau,t,q,p;\hbar)$ are the
corresponding elements in the phase space after  performing the WWM map. The
main problem with this approach is that we do $not$ have representations of the
continuum {\bf Z}-graded Lie algebras in the Hilbert space, $L^2(R^1)$ and,
consequently, we cannot evaluate the matrix elements $<q-{\xi\over 2}|{\hat L}(r,\tau)|q+{\xi \over
2}>;<{\hat H}>$.  
For this reason we have to recur  to another methods to solve this problem.  

The quantities ${\cal L},{\cal H}$ are defined as  :

$${\cal L}(\tau,t,q,p;\hbar )\equiv \int ^\infty_{-\infty}~d\xi <q-{\xi\over
2}|{\hat  L}(\tau,t)|q+{\xi \over 2}> exp [{i\xi p\over \hbar}].
\eqno (5)$$

 $${\cal H}(\tau,t,q,p;\hbar )\equiv \int ^\infty_{-\infty}~d\xi <q-{\xi\over
2}|{\hat  H}(\tau,t)|q+{\xi \over 2}> exp [{i\xi p\over \hbar}].
\eqno (6)$$
the latter matrix elements, if known,  suffice to determine the 
quantities  ${\cal L}$ and ${\cal H}$  associated with 
the $2D$ continuous Toda molecule equation.

Despite not knowing the explicit operator form of ${\hat L}(\tau,t)$ and ${\hat
H}(\tau,t)$ acting in the Hilbert space, 
$L^2(R^1)$, one may still write down solutions for the continuous Moyal-Toda 
equation. This can be achieved starting from the original Moyal SDYM/ASDYM
equations 
associated with the $SU(\infty)$ group in $D=4$ and looking for solutions. 
Ivanova and Popov [3] , in a summary of YM equations in $D\ge 4$, have
discussed  that solutions to the ASDYM/SDYM equations in 
$D=4$ for an arbitrary Lie group, $G$,  which are $linked$  to the Nahm equations 
may be obtained from the ansatz :

$$A_\mu =-\eta^\alpha_{\mu\nu} T_\alpha (\phi (x^\mu))\partial_\nu \phi.~~~
\eta^\alpha_{\mu\nu} =\epsilon^\alpha_{\beta \gamma}.~~~\eta^\alpha_{\mu 0}=
-\eta^\alpha_{0\mu }=\delta^\alpha_\mu. \eqno (7)$$
The 'T Hooft matrices obey the quaternionic algebra :

$\eta^\alpha_{\mu\lambda}\eta^\beta_{\lambda\nu}=-\delta^{\alpha\beta}
\delta_{\mu\nu}-
\epsilon^{\alpha\beta\gamma}\eta^\gamma_{\mu\nu}.$
The function $\phi$ obeys :$\partial_\mu \partial^\mu \phi =0$ (ASDYM ) and $\phi
=x_\mu x^\mu$ (SDYM )
and the Lie algebra valued $functions$ $T_\alpha (\phi) =T^A_\alpha (\phi) L_A$,
for $\alpha =1,2,3$ satisfy
the Nahm quations w.r.t the $\phi$ function :

$$\epsilon _{\alpha\beta\gamma}{dT_\gamma \over d\phi}=\pm [T_\alpha,
T_\beta].\eqno (8)$$
where the $\pm$ corresponds to the SDYM/ASDYM case. Notice that the simple
reflection : $T_\alpha \rightarrow -T_\alpha$ converts the SDYM to the ASDYM
solutions with the proviso that now $\phi$ obeys the $4D$ Laplace equation. It is very important to emphasize that Ivanova and Popov are using an Euclidean spacetime signature . This will beome important later on when we discuss other results obtained in a $(++,--)$ signature.  

The Ivanova-Popov ansatz, for Euclidean signatures,  will yield solutions
to the Moyal deformations of the Anti-Self Dual Gravitational equations in $4D$ 
from dimensional reductions of the $SU(\infty)$ ASDYM equations. However, this
will not be the case for the self-dual gravitational equations that can be 
obtained from reductions of the $SU(\infty)$
SDYM equations. The Ivanova-Popov ansatz will not yield solutions to the Moyal
deformations of the SDG equations. Another type of solutions will be required. For signatures, $2+2$ the situation is reversed. More on this issue will be clarified in the next sections.

A WWM quantization requires writing down the symbol map of  the 
$operators$ acting in the Hilbert space, $L^2 (R^1)$, associated with the three 
Lie algebra valued functions, $\hat T_\alpha$ so the Moyal Nahm equations are :

$$\epsilon _{\alpha\beta\gamma}{d{\cal T}_\gamma \over d\phi}=
\pm \{ {\cal T}_\alpha,{\cal T}_\beta \}_{Moyal}.\eqno (9)$$
with 
${\cal T}_\alpha (\phi (x^\mu;q,p,\hbar);q,p,\hbar) =symbol~ [{\hat T }_\alpha]$ where
${\hat T }_\alpha$ is a
representation of the Lie algebra valued operators in $L^2 (R^1)$. 

Rigorously speaking one should write ${\cal T}_\alpha [{\cal G} ]$ to include the explicit dependence on the Lie algebra  ${\cal G} $ involved initially in the construction. 
In the case that $G=SU(\infty)$ one is required then to $extend$  the symplectic diffs in $2D$ to $4D$. The Moyal bracket involves now a generalized phase space $q_i, p_i$ for $i=1,2$. For $G=SU(2)$ the fact that the dual of the $SU(2)$ Lie algebra is $R^3$ allows to view the three scalars ${\cal T}_\alpha$ as the three components $X,Y,Z$ of a four vector, after fixing the gauge $A_0=0$, replacing $\phi $ by $\tau$ and making the correspondence $A_x \leftrightarrow X, ...$.

Furthermore, if Moyal deformations of the Toda equations are related to the Moyal Nahm equations via the Lax pair formalism, the scalar $\phi (x^\mu)$ will require in general a deformation of the type : $\phi (x^\mu;q_i,p_i,\hbar)$  obeying a deformation of the Laplace equation. The latter equation is in general modified to include $\hbar$ corrections and derivatives acting on the phase space coordinates as well. 
This can be explicitly seen when one performs a direct Moyal quantization program of all the equations involved in the Ivanova-Popov construction. Operators are mapped into functions of phase space via the symbol map. Products of operators are mapped into the Moyal star product of their corresponding symbols and this involves derivatives of arbitary order w.r.t $q,p$. Furthermore, a suitable ordering prescription must be specified a priori. Therefore, a Moyal quantization of the Ivanov-Popov construction induces a $q.p,\hbar$ dependence on the scalar $\phi$ and it deforms the original Laplace equation.   
We shall go back to this issue when we discuss the twistor transform mapping the nonlinear $3D$ Toda equation  
into the $3D$ Laplace equation for the scalar $\phi$.

A particular class of solutions of the $SU(2)$ Moyal-Nahm equations in terms of
the Jacobi elliptic functions w.r.t the undeformed $\phi$  has been
given by [10] :

$${ \cal T}_1 =sn [\phi]({i\over 2}p(q^2-1)-\hbar (\beta +{1\over 2})q].~~~
{ \cal T}_2 =dn [\phi](-{1\over 2}p(q^2+1)-i\hbar (\beta +{1\over 2})q].$$

$${ \cal T}_3 =cn [\phi](-ipq-\hbar (\beta +{1\over 2})]. ~~~\beta=constant\eqno (10)$$
where the ( undeformed) scalar function $\phi$ has a correspondence with $one$,  and only one,  temporal
parameter , $\phi \rightarrow \tau$  ( which is clearly a solution of Laplace equation in the
ASDYM case but does not correspond to the $\phi =x^\mu x_\mu$ required 
in the SDYM case). When $\phi\rightarrow \tau$ the ansatz of
(7) gives that $A_0=0, A_i \sim {\cal T} _i $ and  as expected the $SU(2)$ Moyal
Nahm equations involve the three  components of a $four$ vector and the
derivatives are taken w.r.t the temporal variable $\tau$ that does $not$
transform as a scalar like $\phi$. It is important also to remark that if $\phi $ acquires a deformation $\phi (x^\mu; q,p,\hbar)$ the solution given in (10) will $no$ longer hold !

As stated earlier, by $SU(2)$ Moyal Nahm it is meant that the
solutions (10) in [10] were obtained by performing the WWM map 
 which takes $su(2)$ Lie algebra-valued operators
belonging to the Hilbert space, $L^2(R^1)$ , into functions of $q,p$. 
Beforehand, the Wolf representation of
the $su(2)$ Lie-algebra valued generators, in matrix form using the Pauli spin
matrices, $A_i^a\tau_a$, needs to be used prior to evaluating the WWM map. i.e;
it was essential to use a representation which takes the three Pauli spin 
$SU(2)$ matrices into three known 
operators in ${\hat q}, {\hat p}$.  It is in this sense that one may speak of the solutions (10) to the 
$SU(2)$ Moyal Nahm equations.

Since representations of $SU(\infty)$ in terms of operators in the Hilbert space $L^2 (R)$ are not known ( as far as we know) one cannot evaluate explicitly the WWM map. In addition, $SU(\infty)$ requires to use the $extended$ phase space which implies that the Moyal bracket to be used in (9) will be the one w.r.t the $q_i,p_i$ phase space coordinates rather than to $q,p$. We have symplectic diffs in four dimensions instead of two-dimensions. 
Therefore, the  generalized Moyal Nahm equations require an $extra$ set of variables : $q',p'$ 
that must be introduced to account for the area-preserving diffs algebra
associated with a $4D$ manifold instead of a two-dim surface ( sphere, torus) so now ${\cal T}_\alpha
(\phi;q,p,\hbar;q',p')$
The Moyal brackets are then computed w.r.t the $enlarged$ phase space involving
the $q,p$ and $q',p'$. We shall discuss this in detail in the last section.  

Nevertheless, the generalized "$SU(\infty)$"
Moyal-Nahm equations admit $reductions$ to the continuous Moyal Toda equations
and equations directly linked to the $4D~SU(\infty)$ Moyal ASDYM/SDYM equations.
We refrain from using the term $SU(\infty)$ Moyal Nahm because it is $not$ 
really a WWM quantization of the classical $SU(\infty)$ Nahm equations but , instead, 
one has Moyal
deformations of the algebra of symplectic diffs in $4D$. Such algebra is an
infinite dimensional extension of the area-preserving diffs of a two-dim
surface [20].

The reduction one is referring  can be attained  
simply by imposing the dimensional reduction conditions : $q=q', p=p'$ which
will reduce the phase space
in half and the Moyal bracket will involve one pair of phase space variables
only and $not$ two
copies of the same pair.
In this
fashion one has now a set of functions depending on $six$ variables  instead of
$eight$ :

$$ {\cal T}_\alpha (\phi (x^0,x^1,x^2,x^3);q,p,\hbar). \eqno (11a)$$

From now on, by $reduction$ of the generalized  Moyal Nahm equations we mean
that reduction implied by eq-(11a).  A futher dimensional reduction/restriction on the $x^\mu$
coordinates is needed in order to match the
$t,\tau$
variables required in the original classical continuous Toda equation, 
$\phi$ must be  after a dimensional
reduction $\phi =\phi (\tau,t)$. We are omitting for convenience the $q,p,\hbar$ dependence on the scalar $\phi$.

 Firstly, this may 
obtained by a dimensional reduction  of the six variables to four in (11a); i.e
from $6D$ to $4D$, which is how in the first place the
Moyal SDG  equations are obtained from a dimensional reduction of 
$4D~SU(\infty)$ SDYM [7,8]. A further rotational Killing symmetry reduction 
is needed from the
Moyal heavenly  equation to the $3D$ continuous Toda theory . First of all, in
the ASDYM case, the undeformed $\phi$ obeys the $4D$ Laplace equation $\partial_\mu \partial^\mu
\phi =0$ .  Upon complexification one may choose the variables $x^\mu$ to be 
related to $four$ out of the six variables : 

$y,z,{\tilde y},{\tilde z};q,p,$ and the dimensional reduction is chosen for 
the ( undeformed) scalar $\phi$ as follows :

$$\phi (x^\mu (y,z,{\tilde y}, {\tilde z}))
\equiv\phi (w,{\tilde w}). ~~~ w=z+{\tilde y};~{\tilde w} ={\tilde z}-{ y}.
\eqno (11b)$$
so the functions are of the form  ${\cal T}_\alpha [\phi (w,{\tilde w});q,p,\hbar]$.

The dimensional reduction conditions in  [8,9] were of the type :

$$w=z+{\tilde y}.~~~\partial_w=\partial_z =\partial_{{\tilde y}}. ~~~                            
{\tilde w} ={\tilde z} -{ y}. ~~~\partial_{{\tilde w}}=\partial_{{\tilde z}}
=-\partial_{{y}}. \eqno (12a )$$                             
with 

$${\tilde y} =x^2-ix^3.~~~{\tilde z}=x^0+ix^1.~~~y=x^2+ix^3.~~~~
z=x^0-ix^1.\eqno (12b)$$
The ASDYM equations in $4D$ Euclidean spacetime read : $F_{y{\tilde y}}+F_{z {\tilde z}}=0$ and $F_{y{z}}=0$ , 
$F_{{\tilde y} {\tilde z}}=0$. For signature $(++,--)$ the situation is reversed, the SDYM equations are the ones of the form  $F_{y{\tilde y}}+F_{z {\tilde z}}=0$ instead. The Moyal  heavenly form is : 
$\Omega (w,{\tilde w}, q,p,\hbar)$  and in
effect a dimensional reduction from a $6D$ theory to a $4D$ is
obtained. The Toda requires an additional reduction that shall be discused
below. The Moyal deformed YM potentials are  [8,9]:

$$A_{{\tilde y}}=\partial _{{w}}\Omega -{1\over 2} {\tilde z}.
~~~A_{{\tilde z}}= \partial_ {{ \tilde w}}\Omega +{1\over 2} {\tilde y}. 
~SDYM.~~~(++,--)~signature. \eqno (13a) $$
Under ${\cal T} \rightarrow -{\cal T}$, then $A\rightarrow -A$ and the ASDYM
case is retrieved :

$$A_{{\tilde y}}=-\partial _{{w}}\Omega +{1\over 2} {\tilde z}.
~~~A_{{\tilde z}}= -\partial_ {{ \tilde w}}\Omega -{1\over 2} {\tilde y}. 
~ASDYM.~~~(++,--)~signature. \eqno (13b) $$
so that the Moyal heavenly equations are equivalent to a zero curvature
condition :

$$ \{ \Omega_{,w}, \Omega_{,{\tilde w}}\}=\pm 1 \leftrightarrow 
\{ \partial _{{\tilde z}} +A_{{\tilde z}}, \partial_{{\tilde y}} +A_{{\tilde y}}
\}=0. \eqno (14)$$ 
The minus sign , $-1$ is assigned to the ASDYM case 
related to Anti Self Dual Gravity in $4D$ with $2+2$ signature. Whereas, the $+1$ sign is the one corresponding to the SDYM case with $2+2$ signature related to Self Dual Gravity.  In the Euclidean signature case, the signs appearing in front of the $1$ coefficient are reversed. The important thing is that the (undeformed) scalar obeying  Laplace equation is the solution that must be used 
for Self Dual Gravity in $2+2$ . This coresponds precisely to the ASDYM equations in $(4,0)$ given by Ivanova and Popov.

Therefore, one learns from eqs-(7,13) 
that there is a direct relation between the three 
functions ${\cal T}_\alpha$ appearing in (7) and the $\Omega$ obeying (14) using
the reductions (12). Before this can be achieved it is essential to perform a
gauge transformation from the YM potentials in (13) to the new ones  :

$$ A_\mu \rightarrow A_\mu +\partial_\mu \Lambda
+\{A_\mu,\Lambda \}. ~~~\delta F_{\mu\nu} =\{F_{\mu\nu}, \Lambda\}. $$
Choosing a $\Lambda$ depending on $y,z,{\tilde y}, {\tilde z}$ only and not on
$q,p$ gives :

 $${\hat A}_{{\tilde y}} =A_{{\tilde y}}+\partial_{{\tilde y}} \Lambda
(y,{\tilde y},z,{\tilde z}).~~~
{\hat A}_{{\tilde z}} =A_{{\tilde z}}+\partial_{{\tilde z}} \Lambda
(y,{\tilde y},z,{\tilde z}). ~~~F'_{\mu\nu} =F_{\mu\nu}=0 \eqno (15a) $$
The choice $\Lambda ={1\over 2} (y{\tilde y} +z{\tilde z})+f(y,z)$ is the
adequate one that yields  ${\hat A}_{{\tilde y}}, {\hat A}_{{\tilde z}}$ as
functions solely of $w,{\tilde w},q,p$.  
It is now that one can equate in the ASDYM case  :

$$\partial_{{\tilde z}} {\hat A}_{{\tilde y}}=\partial_{{\tilde w}} 
{\hat A}_{{\tilde y}}=
\partial _w\partial_ {{\tilde w}}\Omega +{1\over 2} .$$

$$\partial_{{\tilde y}} {\hat A} _{{\tilde z}}= \partial_{w} {\hat A}_{{\tilde z}}=
\partial_{{\tilde w}}\partial_ {{ w}}\Omega -{1\over 2}.  \eqno (15b) $$         
where the YM potentials 
${\hat A} _{{\tilde y}}={\hat A} _2-i{\hat A} _3,$ and 
${\hat A} _{{\tilde z}}={\hat A} _0+i{\hat A} _1$ obtained from 
the ansatz (7), are  :  
$${\hat A}_{{\tilde y}}=\eta^\alpha_{2\mu}T_{\alpha}\partial_\mu \phi -i
\eta^\alpha_{3\mu}T_{\alpha}\partial_\mu \phi . \eqno (16a)$$

$$ {\hat A}_{{\tilde z}}=
\eta^\alpha_{0\mu}T_{\alpha}\partial_\mu \phi +i
\eta^\alpha_{1\mu}T_{\alpha}\partial_\mu \phi. \eqno (16b)$$

Concluding, plugging the values of the YM potentials (16) directly into (15) yields the
solutions of the Moyal  ASDG equations (14) related to the $reductions$
of the  
generalized  Moyal Nahm equations ( when the phase space is reduced in half,
$q=q',p=p'$)  that are $encoded$ in the ansatz (7) which
solves the ASDYM equations in $4D$ for the $SU(\infty)$ . The Moyal 
heavenly form is then, up to integration "constants", $f(q,p)$ :

$$\Omega (w,{\tilde w},q,p,\hbar) =\int({\hat A}_{{\tilde y}}-{1\over 2} {\tilde w})dw + 
\int({\hat A}_{{\tilde z}}+{1\over 2} {w})d{\tilde w} . \eqno (16c)$$ 
where the YM potentials are explicitly given by ( 16a,16b). Eq-(16c) is the main
result of this section. It will be shown later on that eqs-(16), via the Lax pair formalism described in eq-(19) , contain the map taking the Moyal Heavenly form into the $SU(2)$ Moyal Toda field obeying the Moyal Toda equations described in section {\bf III}.  

It remains to
determine the ( undeformed) function $\phi$, whereas  the ${\cal T}_\alpha$ may be solved as follows
, for example,  
by using  the solutions of the  $SU (2) $  
Moyal-Nahm equations given by (10) that require the ( undeformed )   scalar $\phi$ without explicit $q,p,\hbar$ dependence.

$$-{ \cal T}_1 =sn [\phi (\tau,t)][({i\over 2}p(q^2-1)-\hbar (\beta +{1\over2})q].~~~
-{ \cal T}_2 =dn [\phi (\tau,t)][(-{1\over 2}p(q^2+1)-i\hbar (\beta +{1\over
2})q].$$

$$-{ \cal T}_3 =cn [\phi (\tau,t)][(-ipq-\hbar (\beta +{1\over 2})]. \eqno (17)$$
The minus signs are due to the fact that the ASDYM equations are the ones
related to the solutions of the Laplace equation and where it is $essential$
that this embedding restricts the space of solutions 
$\phi (x^\mu)$ to depend solely on $\tau,t$. One must use the undeformed $\phi$. An explicit example will be given
shortly. 
In this fashion, the direct relation between  the Moyal deformed 
ASDYM potentials appearing in (16,17) and  the Moyal  heavenly form 
obeying (14)  may be  obtained from eqs- (15).

These reductions are also compatible with the fact that 
the Toda equations for $SU(N)$ are obtained from particular reductions of
Nahm equations which, in turn, can be represented in a Lax pair form :

$$L=T_1 (\phi) +iT_2 (\phi).~~~iT_3 (\phi) =M.~~~{dL \over d\phi }=[L,M]. 
\eqno (18)$$
Consequently ,  in the $N=\infty$ limit,  
one can recast the continuous Moyal-Toda equations  in the double commutator
form after establishing the following
$correspondence$ ( which are $not$ identifications)   , see (4),   :

$${\cal L}(u,\rho)\leftrightarrow {\cal T}_1 +i{\cal T}_2.~~~i{\cal T}_3 ={\cal M} \leftrightarrow 
\{ {\cal L}, {\cal H}\}.~~~{\partial {\cal L} \over \partial \tau}=
\{ {\cal L}, \{ {\cal L}, {\cal H}\}\}                   \eqno (19)$$
where the ${\cal T}_\alpha$ obey (9). 
Setting aside at the moment the differences between the SDYM and ASDYM cases ( the undeformed $\phi$ differs in each case) one can see that eqs-(16,19) contain the relationship between the Moyal heavenly form, $\Omega$ and the Toda field $\rho$ via the YM potentials which, in turn, are expressed in terms of the 
Moyal-Nahm functions ${\cal T}_\alpha$ by using eqs-(16a,16b).

Therefore, barring the differences between the SDYM and ASDYM cases, eqs-(16,19) express the indirect mapping bewteen the Moyal heavenly form and the Toda field via the Lax pair formalism. The main problem is to find an $explict$ realization of the ${\cal L}, {\cal M} $ functions which amounts,  again, to finding a representation of the continuum {\bf Z} graded Lie algebras in the Hilbert space $L^2( R)$.

If one makes  a strict  identification  in eqs-(19) ,  instead
of a correspondence,  from eqs-(9,18,19) one learns that
( the undeformed) $\phi$ must be restricted further.  It then obeys   the  additional condition   :

$${\partial {\tilde \phi } \over \partial \tau }={\partial \phi \over \partial x^\mu}
{\partial x^\mu \over \partial \tau}=
{\partial {\hat \phi} \over \partial w}
{\partial w\over \partial \tau}+  {\partial {\hat \phi } \over \partial {\tilde w}}
{\partial {\tilde w}\over \partial \tau}=1 . \eqno (20)$$
with   $w, {\tilde w} $ functions of $\tau,t$ . Eq-(20) amounts to constraining one out of the four coordinates $x^\mu$.

 We will show at the end of this section 
that   
this additional restriction  on ( the undeformed) $\phi$ would correspond to a very special class of solutions  to the
$\rho$ where the $\tau,t$ dependence of the $\rho$ is required to be of the form : $\rho (\tau,t) =
\rho (\tau \pm it)$.   Therefore, identifying the l.h.s of (19) with the r.h.s furnishes a  restricted
class of solutions to the Moyal deformations of the Lax-Brockett equations (4)  given in terms of  
special solutions to the Moyal-Nahm equations.  The restriction is due to the fact that $\phi$  is
constrained to be  $ \tau \pm it$.   This restriction can  be avoided  by establishing  the
correspondence  given by   eq-(19).

The reason that 
one can make the  correspondences ( which are $not$ identifications)  given by (19) is because there
are  continuum Lie algebras that are isomorphic to Poisson bracket algebras $\sim su(\infty)$ , 
which correspond to the Lie algebras of  are-preserving diffs of the sphere, torus [13]. 
It is in this sense that  the correspondence of eq-(19) is implemented.   There two ways to retrieve
Moyal Toda equations : one way is to use the Lax-Brockett double commutator form and another to use the
Lax representations for the $SU(\infty)$ Nahm equations.   The correspondence between these two 
constructions  of  the Moyal  equations originates  from the fact that   there is a Legendre -like
transform that maps solutions  of   the  $2+1$ continuous Toda equation to those of the three-dim
Laplace equation. i.e. $\rho \rightarrow \phi$.

The $2+1$ continuous Toda equation occurs  in the theory of self-dual Einstein spaces and has a
 well known Eguchi-Hanson solution [ 26] . 
Prasad  [25] ( and discussed by Chapline in [5]) has shown that by a change of variables one can transform the Toda equation into a three
dimensional Laplace equation for a certain function $ \phi =\phi (\rho)$ related to the two-center 
gravitational instanton of Gibbons and Hawking [26] .    
So the correspondence dictated by eq-(19) is a reflection of  the Legendre-like map which takes the
Toda field  $\rho$ to  the $\phi$ obeying  the Laplace equation after a dimensional reduction from
$2+1$ to $2$ dimensions.  The Legendre-like transform determines the correspondence given in (19),
once the suitable maps from the remaining $u, {\kappa}$ functions to $\phi$ are found.

As stated earlier, a  subtlety will now  arise. Due to the Moyal deformations of the Toda equations it is expected then that an accompanying $deformation$ of the Laplace equation for the scalar field $\phi$ follows. The Prasad map taking $\phi \rightarrow \rho$ must be deformed as well. Therefore one should be forced to include $\hbar$ corrections to the scalar $\phi (x;\hbar)$. Similar considerations have been found by [27] 
$after$ employing the WWGM map : an explicit introduction of $\hbar$  was made afterwards. Nevertheless, the correspondence given by eq-(19) still holds once the deformed map from $\phi$ ( obeying deformations of Laplace equation)  to $\rho $ ( obeying deformations of the Toda equation) is found. Similar arguments apply to remaining functions : $u, \kappa$. These also acquire an $\hbar$ 
explicit dependence. 

In general, there must be an explicit dependence on the $q,p$ phase space coordinates for the continuous Moyal Toda field $\rho (\tau, t; q,p,\hbar)$. This is implies that the scalar $\phi=\phi (x^\mu; q,p,\hbar)$ as mentioned earlier. However, in the case of the $SU(2)$ Moyal Toda field, which is really a Toda lattice theory whose discrete spacing is a multiple of $\hbar$, one has instead 
that $\psi (\tau,t;\hbar)$ in agreement with [16,27]. 
The latter case implies then that the scalar $\phi=\phi (x^\mu;\hbar)$ only.  
This will become more transparent in the course of the text. 

There is also another reduction from the generalized Moyal Nahm equations to  
Moyal Toda equations which shall be studied in {\bf IV} that differs from the reduction
obtained from eq-(11a). Such reduction involves using the enlarged phase space variables $q,p,q',p'$;  ( using the full symplectic diffs in four dimensions). 
It is obtained by setting ( the undeformed )  $\phi\rightarrow \tau$, $t=q'$ and by integrating out the $p'$ variable
after finding  the three functions ${\cal T}_\alpha$ which  solve the generalized Moyal-Nahm
equations.  Whereas the reduction obtained from (11a) involves incorporating the $\tau,t$ variables
through the ( undeformed) scalar function $\phi (\tau,t)$ and establishing  the correspondence with  Lax-Brockett
formalism. Both types of reductions lead to Moyal Toda equations.

The  very special case 
we are studying in this section  requires that one performs a series of coordinate redefinitions and
dimensional reductions from $6D$ to $4D$ :

$$\{x^0,x^1,x^2,x^3 ;q,p\} \rightarrow \{y,z,{\tilde y}, {\tilde z};q,p\}  
\rightarrow \{w, {\tilde w}; q,p\}
\rightarrow \{\tau,t; q,p \}. \eqno(21a)$$
The Laplace equation 
fixes  the family of ( undeformed) functions $\phi$  . 
Due to the dimensional reduction, the $4D$ Laplace operator $degenerates$ to zero
, this can be verified by simple inspection :

$$w=x^0-ix^1+x^2-ix^3.~~~{\tilde w} =x^0+ix^1-x^2-ix^3.~~~w^*\not={\tilde w}. $$

$$\partial_0=\partial_w +\partial_{{\tilde w}}.~~\partial_1=-i\partial_w
+i\partial_{{\tilde w}}.~~\partial_2=\partial_w-\partial_{{\tilde w}}.~~
\partial_3 =-i\partial_w -i\partial_{{\tilde w}}. \eqno (21b)$$

One can verify that the Laplace  operator acting on ( the undeformed) $\phi$ : 

$$\partial^2_0+\partial^2_1+\partial^2_2
+\partial^2_3=\partial_y \partial_{{\tilde y}} +\partial_z \partial_{{\tilde z}}=
\partial_w \partial_{{\tilde w}} - \partial_w \partial_{{\tilde w}} 
\equiv 0 \eqno (21c)$$
hence, as a result of the dimensional reduction, $\phi (w,{\tilde w})$, the $4D$ Laplace
operator acting on ( the undeformed) $\phi$ degenerates to zero. i.e. $Any$ function of the 
form
$\phi =\phi (w,{\tilde w})$ $obeys$ automatically the $4D$ Laplace equation. 

If one wishes one may restrict the solutions of the $4D$ Laplace equation,
$\partial_\mu \partial^\mu \phi(w,{\tilde w})=0$, for arbitary functions $\phi$ 
to those obeying the 
$2D$ Laplace equation , instead, :  

$$\partial^2_\tau \phi +\partial^2_{t}\phi = \partial_w \partial_{{\tilde w}}
\phi =0 \Rightarrow \phi= f(w)+g({\tilde w}).~~~w\equiv \tau+it.~~~{\tilde w} =\tau-it. \eqno (21d)$$
In this fashion one can remove the arbitrariness of $\phi$. 

It is important to emphasize that ${\tilde w} \not= w^*$ and that $\tau,t$ are
$complex$ valued since the Moyal heavenly  equations are defined in 
$complexified$ $4D$ spacetime. A real slice may be taken by choosing 
${\tilde w} =w^*$ which implies that $\tau,t$ must be real. 
The  general solution to the $2D$ ( complexified in general )  ( undeformed) Laplace equation for the (undeformed) $\phi$ is :
$\phi=f(\tau+it)+g(\tau-it)$ for $f,g$ arbitary. Hence, one may choose without
loss of generality the following solutions to the $reductions$ of the 
generalized Moyal-Nahm
equations related to the $SU(\infty)$ ASDYM equations in $4D$ :

$$-{ \cal T}_1 =sn [f+g][({i\over 2}p(q^2-1)-\hbar (\beta +{1\over 2})q].~~~
-{ \cal T}_2 =dn [f+g][(-{1\over 2}p(q^2+1)-i\hbar (\beta +{1\over 2})q].$$

$$-{ \cal T}_3 =cn [f+g][(-ipq-\hbar (\beta +{1\over 2})]. 
~~~\phi=f(\tau+it)+g(\tau-it).\eqno (22a)$$
Inserting the above solutions into eqs-(16a-16b) give automatically through
eqs-(15b) an explicit
solution, eq-(16c),  to the Moyal deformations of the ASDG equations in $4D$
given in eq-(14).  
The above solutions to the $reductions$ of the  generalized Moyal-Nahm equations are parametrized by
an arbitrary family of functions of right/left movers, $f (\tau+it), g(\tau-it)$.

At this point one is ready to examine the meaning of the correspondence of eq-(19).  If an strict
identification is made, 
eq- (20)  will restrict these family of right/left movers 
$f,g$ to satisfy the following additional constraint in 
$addition$  to the two-dim Laplace equation :

$$ {\partial \phi \over \partial \tau}=1~and~\partial^2_\tau \phi +\partial^2_t \phi =0 
\Rightarrow \phi= \tau+at+b  =f+g. $$
There are three simple cases to consider , take $b=0$,  :

$$(i)~f+g=\tau \Rightarrow a=b=0.~~~(ii) f+g =\tau+it \Rightarrow g=0.~~~(iii)
f+g=\tau-it \Rightarrow f=0. \eqno (22b)$$ 
case (i) belongs to    $f(\tau+it)\equiv {1\over 2}(\tau+it), g(\tau-it)\equiv
{1\over 2} (\tau-it)$ so $f+g=\tau$.  This case bears a direct connection to the $SU(2)$ Moyal Nahm
equation found by Strachan [16], eq-(29) in the next section.
 Cases (ii), (iii), are typical of the above mentioned
restricted solutions  of the type
$\rho(\tau,t)=\rho (\tau\pm it)$ that bear a connection to the Moyal Toda equation of the type given
by eq-(26) in the next section.  
These are  very special cases. In general ,  the scalar field
$\phi =f+g$ for arbitrary $f,g$  so one should focus on the most general solution for $\phi$ instead
of a very restricted special case.

The
main point of these results is that there $must$ be an additional dependence on
$t$ ( besides the original $\tau$ dependence ) in the formulation of the
$reductions$ of the 
generalized Moyal-Nahm equations. Having found a particular choice 
for ( the undeformed) $\phi =f(\tau+it)+g(\tau-it)$, which $solely$ depends on $\tau,t$,  
 a solution to
eqs-(9) like eqs-(17,22a) will finally yield the sought after solution to the 
$4D~SU(\infty)$
ASDYM  equations via the
ansatz (7), as it has been explicitly provided above by eqs-(14,15,16,17). Once
a solution has been found other solutions may be attained by the action of the
infinite dimensional group of symmetries of the anti/self dual $SU(\infty)$ 
Yang-Mills and heavenly equations [17]. The symmetry group is an infinite
dimensional affine Kac-Moody algebra associated with the algebra of
area-preserving diffs.

The SDYM equations require a separate study since in this
case ( the undeformed) $\phi =y{\tilde y} +z{\tilde z}$ is fixed and does not obey the Laplace
equation. A different kind of reduction other than the Ivanova-Popov ansatz, via
the Moyal Nahm equations,  is necessary to obtain the Moyal
SDG equations from the $4D~SU(\infty)$ SDYM equations. See [7,10].  By setting :

$$\phi =x^\mu x_\mu=y{\tilde y} +z{\tilde z}=u+v.~~~u\equiv y{\tilde y}.~~~v\equiv 
z{\tilde z}. \eqno (22c)$$
one has another set of variables besides  the $w,{\tilde w}$  used in
the ASDYM case. But whereas the latter could furnish solutions to the Moyal
deformations of $4D$ ASDG, 
the Ivanova-Popov
ansatz in the former SDYM  case $cannot$ be used to obtain solutions to 
Moyal deformations of $4D$ SDG
although it $does$ yield solutions to the $4D~SU(\infty)$ SDYM equations.  
Notice that one $cannot$ equate the $SO(4)$ invariant quantity $x^\mu x_\mu$
with a non-invariant one like $\tau$ ( a coordinate). Whereas the 
quantity $\phi$ in the
ASDYM case transforms as a scalar , however,  this does not imply that it is $SO(4)$ invariant   
: $\phi (x)=\phi' (x')\not=\phi(x')$.

In  eq-(14) the transformation $\Omega \rightarrow i\Omega$ changes the signs of the r.h.s and  
naively reverses the role of SDYM and ASDYM.  However  eq-( 16c) is $not$ invariant under $\Omega
\rightarrow i\Omega$ : one cannot  absorb the $i$ factors in the $d{\omega}, d{\cal \omega}$  measure
pieces despite being able to absorb them in  each single term inside the integrands. Therefore the naive
analytical  continuation will not convert a SDYM potential into an ASDYM one.  Furthermore, the scalar 
$\phi$ is different in each situation. A mere $i$ multiplication will not suffice.  This fact will
become important when we study the master Legendre transform from  rotationally Killing symmetry
reductions of the Moyal deformations of the  $4D$ Self Dual Gravitational equations  into the Moyal
Toda equations.         

To finalize this section we point out, once again, that the Moyal Toda equations obtained     
from the Moyal Nahm equations  associated with the quantities  ${\cal T}_\alpha [\phi=f+g;q,p,\hbar] $ 
are those related to the $SU(2)$ Moyal Toda equations derived by 
Srachan [16]  ( given by eq-(29) )  and to the continuous Moya Toda equation given by eq-(26) ;  these
equations will be  studied next. 

Concluding, to a first order approximation, neglecting the deformations  to 
the $\phi$ scalar,   
the most general solution to the Moyal  ASDG equations related to the
$reductions$ of the 
generalized Moyal Nahm equations ($q=q',p=p'$)  may be 
parametrized by a family of
undeformed functions obeying the $2D$ Laplace equation 
for the $undeformed$ function $\phi (w,{\tilde w})$ and whose solutions depend on two $arbitary$
functions, $f(\tau+it),g(\tau-it)$ as prescribed in (22a).  

When deformations are included, we can still use eqs-(16,18,19) as the most important equations of this section establishing the indirect correspondence between the Moyal heavenly $\Omega$ form and the Moyal Toda field $\rho$ using the Lax pair formalism. In this case $\phi$ obeys deformations of the Laplace equation obtained by an explicit Moyal quantization of the Ivanova-Popov construction. Hence, it  
contains an additional $q,p$ dependence as it should for the continuous $SU(\infty)$ Moyal Toda field.

In the case of the $SU(2)$ Moyal Toda field ( a 
spatially-discrete but temporal continuous lattice theory ) there is no $q,p$ dependence as it has been shown in [16] and discussed by [27] in the constructions of a master integrable equation that contains the KP, KdV hierarchies....
The explict relation between $\Omega$ and the $SU(2)$ Moyal Toda field will be studied further in {\bf 3.3}.   
We proceed now to study the Moyal Toda equations in the next
section.

\bigskip

\centerline {\bf III.  The Moyal Toda Equations}
\centerline {\bf 3.1. A Continuous Moyal Toda Equation}
\bigskip

In this section we shall display the different forms of the Moyal Toda equations  that are related to the 
the quantities 
${\cal L}(u,\rho), {\cal H} (\kappa)$ in eqs-(5,6,18,19) and the ${\cal T}_\alpha[\phi,q,p,\hbar]$. 
We return now to this discussion.  
After the WWM map is performed ,
$u,\rho,\kappa$ 
$acquire$ and additional 
dependence on $q,p,\hbar$. To illustrate this , let us look at the  
operator form of the original continuous Toda equation :

$${\partial^2 {\hat \rho} \over \partial \tau^2 }={\partial^2 e^{{\hat \rho}}
\over \partial t^2} . \eqno (23)$$

Given an operator, ${\hat \rho}(\tau,t)$,   
acting in the Hilbert space of square integrable functions on the line, of the  form :

$${\hat \rho} =\sum_{mn} \rho_{mn} (\tau,t) ({\hat q}^m {\hat p}^n +{\hat p}^m {\hat q}^n +........) . 
\eqno (24a)$$

with a Weyl ordering prescription  imposed on the monomials in ${\hat q}^m {\hat p}^n$  :

$${\hat q} {\hat p} \rightarrow {\hat q}{\hat p} +{\hat p}{\hat q}.~~~{\hat q} {\hat p}^2  \rightarrow 
{\hat q} {\hat p}^2 +{\hat p}^2 {\hat q} +2{\hat p} {\hat q} {\hat p}.....\eqno (24b)$$
More complicated operators are also possible that  are  not necessary  sums of monomials.  
The WWM map converting operators, ${\hat \rho}(\tau,t)$  into functions in phase
space ( making use of the
symbol map ) yields  :

$$symbol~ [{\hat \rho}] =\rho (t,\tau,q,p,\hbar).
~symbol~ [e^{{\hat \rho}}] =e^{*\rho}=1 +\rho +{\rho*\rho \over 2!}
+{\rho*\rho*\rho \over 3!} +........ \eqno (25)$$

Hence the putative Moyal continuous Toda molecule equations reads :

$${\partial^2  \rho \over \partial \tau^2 }={\partial^2 e^{{*\rho}}
\over \partial t^2} . \eqno (26)$$
The Moyal star product of two functions of phase space of dimension $2n$ whose symplectic form has the
inverse  
${\omega}^{IJ}$  is
defined :

$$f*g =\sum _{n=0}^\infty {1\over n!} ({i\hbar \over 2}) ^n  {\omega}^{i_1j_1}
{\omega}^{i_2j_2}.......{\omega}^{i_n j_n}( \partial_{i_1} \partial_{ i_2} ......\partial_{ i_n} f )(
\partial_{j_1} \partial_{ j_2} .....\partial_{ j_n} g). \eqno (27)$$
When ${\omega}^{ij}$ is the inverse of the symplectic form in two-dimensions, the derivatives are taken 
w.r.t the $q,p$ variables only. 
To recover the $2+1$ continuous Toda requires replacing the l.h.s of (26) by  
$(\partial_\tau)^2$  by $\partial_{z_+}\partial_{z_-}$ and setting 
$\rho (z_+, z_-, t,q,p,\hbar)$. 
Eq-(26) is one of the Moyal Toda equations equation to be studied in this section {\bf
III}.

The above equation was obtained  from the operator form of the original continuous Toda equation. 
In general, at the quantum level, the form of the operator equations of motion are $not$ the same as
those of the original classical field.  A modification of (26) will be presented shortly where the
r.h.s is modified ; i.e. there will be derivatives of  infinite order w.r.t the $t$ variable that
originate from deformations of the continuum graded Lie algebras as a result of replacing ordinary
Poisson bracket  by Moyal brackets.

The $h=0$ limit of (26) yields in the
r.h.s the ordinary exponential, $e^ \rho$ because
in the classical limit the Moyal star product becomes the ordinary pointwise 
product of functions.  Since in the classical limit, eq-(26) involves a differential equation w.r.t the
variables $\tau,t$ , only,  the classical limit of
(26) does not determine the $q,p$ dependence of $\rho (\tau,t,q,p,\hbar=0)$
which may be completely arbitrary.  Assuming that $\rho$ admits an expansion 
in powers of $\hbar$ : $\rho
=\sum \hbar^n \rho_n (\tau,t,q,p)$ one can impose the condition that the zeroth-order
term does not depend on $q,p$  :
$\rho_0 (\tau,t,q,p)\equiv\rho_{class} (\tau,t).$  In this fashion the $\hbar=0$
limit reproduces the classical continuous Toda equation for $\rho_{class}
(\tau,t)$. This condition can also be derived from the 
master Legendre transform ( studied in {\bf 3.3}) that maps
the rotational Killing symmetry reductions of the Moyal SDG equations into
the continuous Moyal Toda equations [9,21]. Concluding, the zeroth-order term 
must be of the form $\rho_0 =\rho_0 (\tau,t)$.

Another way to obtain the continuous Moyal Toda equations directly  should be  to  perform the master Legendre transform mapping , if indeed it exists, between 
$\Omega (y',{\tilde y}',z',{\tilde z}' ,\hbar)$ obeying the  rotationally Killing symmetry reductions
of the Moyal 
SDG equations  [9,21]  to $\rho
(\tau,t,q,p,\hbar)$ [1,9] obeying the continuous Moyal Toda equations .  
Strachan [16] has shown that  the $SU(2)$ 
Moyal-Nahm equations admit a reduction to the  classical continuous Toda chain
in the $\hbar =0$
limit. Therefore,  reductions of the $SU(2)$ generalized Moyal Nahm equations should yield  the
continuous Moyal Toda equation. This shall be studied in {\bf IV}  .  

We  shall continue shortly with the Strachan ansatz and write down a more general equation than (26)
which contains derivatives of infinite order w.r.t the $t$ variable; i.e.  the operator equations of
motion for the quantized Toda field  $differ$ from the classical counterpart.  
The master Legendre
transform will be discussed  also. The study of the geometry associated with these Moyal deformations
has been
given by [14].

\bigskip

\centerline {\bf 3.2 Strachan's  Reduction  of  the  $SU(2)$ Moyal Nahm Equations}

\centerline {\bf to the $SU(2)$ Moyal Toda Equations}

\smallskip

It is  known that the ordinary continuous Toda equation may be obtained from 
axial-symmetry reductions of the $SU(\infty)$ classical Nahm equations.  This fact permitted Strachan
to construct Moyal deformations  of the Toda equation by replacing the Poisson bracket by the Moyal
bracket. [16] .  If one writes the 
Moyal Nahm equations  of the type given  ( with the plus sign ) by eq-(8) , 
with $\phi=f+g= \tau$  and imposing  the axial-symmetry reductions of the form  :

$$T_1=X_1 =h(\tau,t=q,\hbar)cosp.~~~T_2=X_2 =h(\tau, t=q,\hbar)sinp.~~~T_3=X_3 =z(\tau,t,\hbar). \eqno (28)$$
allows the $decoupling$ of the  $cosp,sinp $ terms  after computing the Moyal
bracket and  , 
after  eliminating the function $z(\tau,t),$.  Strachan [16]  arrived at : 

: 

$${\partial^2 \psi \over \partial \tau^2} ={1\over 4} [{\Delta -\Delta^{-1}\over \hbar}]^2
e^\psi ={\partial^2 \over \partial t^2} e^\psi +{ \hbar^2\over 3}  {\partial ^4\over \partial
t^4}e^\psi + O(\hbar^{2n})..... \eqno ( 29)$$
the shift operators , $\Delta, \Delta^{-1}$ and $\psi(\tau,t,\hbar)$ are defined :

$$\Delta \psi =\psi (t+\hbar). ~~~\Delta ^{-1}\psi =\psi (t-\hbar).~~~e^{\psi/2}
\equiv h(\tau,t). \eqno ( 30)$$
Equation (29) is  the $SU(2)$ Moyal Toda equation. It involves one field only
, $\psi$, in the same way that the Liouville equation is tantamount of a $sl(2)$
Toda field equation.   
In the classical limit, 
$\hbar =0$ one recovers  the $classical$ $continuous$ $SU(\infty)$ Toda
molecule equation as expected. 
This can be  seen by expanding : $\psi (\tau,t,\hbar)
 =\psi_0 +\hbar^2 \psi_2 +.....$ and plugging this value into (29). The
$\hbar=0$ limit reproduces again the classical continuous Toda equation for the field
 $\psi_0 (\tau,t)$ as eq-(26) did for $\rho (\tau,t,q,p,\hbar=0)=\rho_{class}
 (\tau,t).$

It is an interesting question ( although not the right question to ask) if one could find a representation of the $su(2)$
Lie algebra in terms of ${\hat q}, {\hat p}$ operators that would yield
Strachan's solutions after performing the WWM map of the operator equations associated with the $su(2)$
Nahm's equations. The solutions (28) clearly differ from those presented in eq-(10); not only in the
$\tau$ functional dependence implicit in the elliptic functions but also in the dependence of the
phase space variables.  

The reason that one should $not$  view Strachan's construction as a direct WWGM quantization of the $SU(2)$ Toda field, a Liouville theory, is because eqs-(29,30) represent really a 
$SU(2)/SL(2)$ Toda lattice theory whith discrete-spatial spacings in multiples of $\hbar$ : the field $\psi$  is  evaluated at discrete jumps $ t, t\pm \hbar$ , 
. However, there is  continuous temporal dynamics represented by the $\partial_\tau$ derivatives . See the important work of Dimakis et al [28] on this respect
. The r.h.s of (29) is a spatially-discrete-difference operator which can be expanded into an infinite number of derivatives, a nonlocal expression. The Moyal product is also nonlocal due to the infinite number of derivatives. Therefore, loosely speaking one may refer to Strachan's equation as the $SU(2)$ Moyal Toda equation. It has been speculated by Reuter [14] that this maybe the source of nonlocality in Quantum Mechanics.

From the form of (29) one can notice that a WWGM quantization  program  alters the operator form
of the quantum field equations of motion  from the original classical field
equations. The operator equations read instead :

$${\partial^2 {\hat \rho}   \over \partial \tau^2} ={1\over 4} [{\Delta -\Delta^{-1}\over \hbar}]^2
e^{{\hat \rho}} ={\partial^2 \over \partial t^2} e^{{\hat \rho}}  +O(\hbar^{2n})..... \eqno (31)$$ 
the r.h.s involves derivatives of  $infinite$ order  of the operator ${\hat \rho} (\tau,t)$ w.r.t the
$t$ variable . This implies that a WWGM quantization program related to an operator of the type given
by eq-(24) in terms of monomials of ${\hat q}, {\hat p}$,   is directly linked with Moyal deformations
of continuum Lie algebras.  Upon quantization the struture of the Lie algebra is itself modified as
well.  This occurs in the study of quantum groups and quantum Lie algebras where quantum integrability
requires deformations of the classical Lie algebraic structures.  

In view of this, the correct Moyal continuous Toda equation ( it is 
a further deformation of the Strachan's Toda lattice equations by introducing the star exponential )  must be :

$${\partial^2 { \rho}   \over \partial \tau^2} ={1\over 4} [{\Delta -\Delta^{-1}\over \hbar}]^2
e^{{*\rho}} ={\partial^2 \over \partial t^2} e^{{*\rho}}  +O(\hbar^{2n})..... \eqno (32)$$
and is obtained after performing the WWM map of eq-(31).  The Moyal star product is taken w.r.t the
$q,p$ variables only. 
The  extra dependence on the two phase space
variables, $q,p$  is due to the WWM symbol map taking operators into functions
in phase space. 
The $t$ parameter is the one that encodes the continuum Lie algebra generators and commutation
relations.  Upon quantization, the latter algebra is deformed and hence the operator appearing in the 
r.h.s  of the original
Toda equation, $\partial_t^2 $,    acquires a deformation in powers of $\hbar$ 
and an  infinite number of  derivative  terms is induced.  The  deformed
continuum algebra is 
the algebra associated with deformations of  the Poisson bracket algebra of the sphere,  $\sim
SU(\infty)$.  

We have learnt that the $SU(2) $ Moyal Toda equation is $contained$ in the $SU(2)$ Moyal Nahm
equations and, similarily, the $SU(\infty)$ Moyal Toda , eq-(32), 
should  be $contained$ in
the $generalized$  Moyal Nahm equations .  It is precisely for this reason that the
generalized Moyal Nahm equations must depend on an $extra$ set of phase space
variables as argued earlier prior to eq-(11a). This will be study in {\bf IV}
where in particular we shall provide a plausible embedding of (32) into the 
generalized  Moyal Nahm equations.

It becomes important now to study the subtleties due to the choices of different signatures taken by the authors in [3,8]. 
As we stated earlier, the Ivanova Popov ASDYM equations in Euclidean spacetime required the use of the ( undeformed)  scalar field $\phi$ obeying Laplace equation. However, the ASDYM equations in Euclidean space $correspond$ to the SDYM equations in spaces of signature $2+2$ 
studied by [8]. In both cases one has : $F_{y{\tilde y}} +F_{z{\tilde z}}=0$. Reductions of the latter equations are the ones which yield the $2+2$ Self Dual Gravitational equations, and from these, the Toda equations are obtained after a further rotational Killing symmetry reduction. Therefore, in view of this signature subtlety,    
Strachan  solutions (28,29), as expected,   do fit into functions of the type 
: ${\cal T}_\alpha [\phi ( w(\tau,t),{\tilde w} (\tau,t));q,p,\hbar]$ . This can be seen 
by imposing the $reduction$ condition $t=q$ , taking a particular solution for $\phi=f+g$ where 
$f(\tau+it)={1\over 2}(\tau+it)$ , 
$g(\tau-it)={1\over 2}(\tau-it)$
whose sum yields $\phi=f+g=\tau$. And, finally , the ${\cal T}_\alpha$ functions 
have the form  described in eq-(28). 
Concluding, after these steps are taken : Strachan's ansatz is a very special case of the most
general  Ivanova-Popov ansatz for the three functions, ${\cal T}_\alpha$. 

Solutions to (29) may be obtained through iterations after expanding in even
powers of $\hbar$ : $ \psi_o +\hbar^2\psi_2 +....$. In this way an infinite, but known,  number of
differential equations yields iteratively the solutions for $\psi_{2n}$. To solve this system is
another matter.  
The most general solution to the ordinary classical continuous Toda 
equations has been found by Saveliev [13]. It is well known by now that the ordinary Liouville
equation can be embedded into the continuous Toda equations as follows. Given  :

$${\partial^2 \psi_0 \over \partial \tau^2} ={\partial^2 e^{\psi_0} \over
\partial t^2}. \eqno (33)$$ 
One may plug in the ansatz which will automatically reproduce the Liouville
equation :

$$e^{\psi_0} =({1\over 2} t^2+bt+c)e^{\phi_L} \Rightarrow \partial_{z_+}
\partial_{z_-} \phi_L (z_+, z_-) =e^{\phi_L}. \eqno (34)$$
after performing the dimensional reduction $\tau=z_+ +z_-$. 
So, in  the $\hbar=0$ limit eqs-(29) reduces to the ordinary classical continuous Toda molecule 
( chain )
equations. Again, the $2+1$ Toda equations are retrieved by inserting $z_+, z_-$
for $\tau$ and taking $(\partial_\tau)^2$ for $\partial_{z_+} \partial_{z_-}$.
This anstaz ( 34) is the one which reproduces the Eguchi-Hanson  gravitational instanton [26].

\bigskip

\centerline {\bf 3.3 The  Master Legendre Transform}

\bigskip

Before we begin this section we must emphasize that the results of [16,27] are very different from  the results of this section. Mainly because the results of [27] do $not$ involve a rotational Killing symmetry reduction of the Moyal heavenly equations and also because these involve a particular class of dimensional reductions of the Moyal SDYM master equations that do $not$ lead to the Moyal heavenly equations. The authors [27] based their work on the results by Strachan 
[16] on the Toda/KP  hierarchies  and obtained  the  $SU(2)$ Moyal Toda equations from a $different$ route than the one described here. The equations in [16,17] are the $SU(2)$ Toda 
lattice equations whose discrete spacing is a multiple of $\hbar$.   
When $\hbar=0$ one recovers the continuous Toda equations, or the so-called Boyer-Finley equations obtained originally from Killing symmetry reductions of 
the heavenly equations. 

As promised earlier, we can conclude that eqs-(16,19,28) already $contain$  the required map from the $\Omega$ heavenly form obeying the Moyal heavenly equation ,eq-(14),  to the $SU(2)$ Moyal Toda field, $\psi$,  obeying Strachan equation (29). This can be achieved via the Lax pair formalism given by eq-(19) after using Starchan anstaz for the three Moyal-Nahm functions given by eq-(28). This automatically solves the problem of finding the Legendre transform from $\Omega$ to the $SU(2)$ Moyal Toda field 
$\psi$. 

Hence, eqs-(16) yield $\Omega=\Omega [A_i]$. The Strachan solution corresponds to the particular case of $\phi=\tau$ which impies that $A_i=X_i$ and , finally, eqs-(28,29) give the explicit relations $X_i=X_i [\psi (\tau,t,\hbar),z(\tau,t,\hbar)]$ with $exp {\psi\over 2}\equiv h(\tau,t,\hbar)$.   
To study the map from $\Omega$ to the continuous Toda field $\rho(\tau,t,q,p,\hbar)$ is another matter. We proceed to study this case below and show that it can be solved in principle by embedding the $SU(2)$ solutions into the continuous $SU(\infty)$ ones.

It has been known for sometime that rotational Killing symmetry reductions of
the  Self Dual Gravitational or heavenly equations lead to the continuous Toda equations [15]. 
The master Legendre transform , if it exists, between rotational Killing symmetry reductions of 
$\Omega (y',{\tilde y}', z', {\tilde z}',\hbar)$ obeying the $4D$ Moyal SDG equations and $\rho (z_+,
z_,,t,q,p,\hbar)$ obeying the  $2+1$ continuous Moyal Toda equation, eq-(32),  should  be defined as
the  map [9,21] :

$$\Omega (y',{\tilde y}', z', {\tilde
z}',\hbar)\equiv \sum_{n=0}^\infty [{\hbar \over {\tilde y}' (q,p)}]^n 
\Omega_n (r\equiv y'{\tilde y}'; z', {\tilde z}') \rightarrow  
\rho (z_+, z_-,t,q,p,\hbar)=$$

$$\sum_{n=0}^{n=\infty} \sum^{l=n}_{l=0}\sum^{m=+l}_{m=-l}
\hbar^n \rho_{n}^{lm}( z_+, z_- ,t)Y_{lm}(\theta, \varphi). ~~~q=cot (\theta/2)
cos\varphi.~   p=cot(\theta/2) sin \varphi.  \eqno (35)$$
the limits in $l$ are defined with the proviso that in the limit $\hbar =0$ the zeroth order terms will
survive only giving

$$\rho^{00}_0 (\tau,t)Y_{00}(\theta,\varphi) \equiv \rho_{class}(\tau,t). 
~~~lim_{\hbar =0}~e^{*\rho}=e^{\rho_{class}(\tau,t)}. ~~~
lim_{\hbar=0}~\rho=\rho_{class} (\tau,t). \eqno (36)$$
and , as  expected, the zeroth-order terms do $not$ depend on $q,p$.  We have performed also the 
stereographic projection mapping the
sphere into the complex plane ( phase space of $q,p$). This implies that the $n^{th}$ order Legendre
map , if it exists,  must establish the correspondence :

$$ [{1 \over {\tilde y}' (q,p)}]^n \Omega_n (r\equiv y'{\tilde y}';z',{\tilde z}') \rightarrow  
\sum^{l=n}_{l=0}\sum^{m=+l}_{m=-l}
\rho_{n}^{lm}(z_+, z_-, t)Y_{lm}(\theta, \varphi). \eqno (37)$$
an additional dimensional reduction in (33) $z_+,z_-$ to $\tau$ is of course
needed to go from the $2+1$ continuous Toda to the $2D$ Toda molecule equation. 

It is essential not to confuse the prime variables $y',z'...$ with the
variables $y,z...$. A detailed discussion of the maps between the primed and
unprimed variables was given in [9] based on [21] :

$$\{ \Omega_{,z'}, \Omega _{,y'} \}_{{\tilde z}' {\tilde y}'}= 1
\leftrightarrow  
\{ \Omega_{,w}, \Omega _{, {\tilde w}} \}_{ q,p}=1. \eqno (38a)$$ 
The inverse symbol map takes functions of phase space into operators in $L^2(R)$  and Moyal brackets
into commutators leading to the operator equations :

$$   {1\over  i\hbar}   [ {\hat \Omega} _{,z'}, {\hat \Omega} _{,y'} ]= {\hat 1} \leftrightarrow {1\over i\hbar} [ {\hat \Omega} _{,w}, {\hat \Omega} _{, {\tilde w}}  ] ={\hat 1}.\eqno (38b)$$
where 
$$\Omega (w, {\tilde w}, q,p,\hbar) =\sum (\hbar)^n \Omega_n (w, {\tilde w},q,p). \eqno (38c)$$
Extreme care must be taken  $not$  to set $\Omega_n$ as a function of $r=w{\tilde w}$ and $q,p$.  If
this is wrongly  assumed then the r.hs. of  (38) will be $zero$ instead of $1$. This corrects an
errata in [9].  
An  example of  the coordinate transformation ( a field dependent  coordinate transformation up to zeroth order) 
between the primed and unprimed variable  associated
with  the particular solution $\Omega =\Omega_0=z'{\tilde
z}'+y'{\tilde y}'$   
is [9]:

$${\tilde z}'=q.~~~{\tilde y}'=p.~~~z'=w+{\lambda \over q}.~~~y'={\tilde w}
-{\lambda \over p}. \eqno (39)$$
with $\lambda$ a complex constant.  As $n$ runs the coordinate transformation varies  and one speaks of a field dependent 
transformation to the $n$-th order. This is what eqs-(38a,38b) represent.

At first sight matters seem to indicate that one can transform the infinite number of 
differential equations (38) for the $\Omega_n (r\equiv y'{\tilde y}', z', {\tilde z}')$ functions  given
in [9,21] into an infinite number of differential equations for the $\rho^{lm} (z_+, z_-, t)$ functions
after inserting $\rho$ into eq-(32) ( the angular dependence has been factored out). However, there
appears to be  a  discrepancy :  the differential equations for the $\Omega_n$ involve   higher
derivatives w.r.t the
$r, {\tilde z}'$ variables  as the integer $n$ runs from $0$ to $\infty$. Whereas the differential
equations for the $\rho^{lm}_n$ involve higher derivatives w.r.t the $t$ variable  $only$  with solely
quadratic derivatives in $\partial_{z_+} \partial_{z_-}$ appearing in the l.h.s .  

Identical objections arise if one wishes to perform a transformation  from the $\Omega_n$ to the 
$\psi_n (z_+, z_-,t) $ appearing in Strachan's  $SU (2)$ Moyal Toda equation (29) for $\psi$ after
expanding the latter in powers of $\hbar$.  This may  suggest that the l.h.s of eqs-(29,32) should  have
higher derivatives in $z_+, z_-$ or $\tau$ variables if a master Legendre transform exists from the
infinite number of quantities $\Omega_n$ to   the $\rho^{lm}_n$ or $\psi_n$.

Nevertheless this  does not represent a problem for the following reason.  
At any given level $n=N$ , in powers of $\hbar$ ,  the number of differential equations involve solely 
$\Omega_0 , \Omega_1,......\Omega_N$  . The number of derivatives w.r.t the $r, {\tilde z}'$  is higher
for the $\Omega_0$ function than for the $\Omega_N$ .  The  number of derivatives w.r.t the $r, {\tilde
z}'$  at level $n=N$  is the lowest for the $\Omega_N$ function because $\Omega_N$ appears  already
with a power of $\hbar^N$ .  Hence , the number of derivatives  of  $\Omega_N$ w.r.t   the $r, {\tilde
z}'$ variables   precisely matches the number of $\tau$ or $z_+, z_-$ derivatives acting on
$\rho^{lm}_n$  [9,21] at that given order $n=N$. Starting from the zeroth order term,  by iteration ,
one can span the whole range of $\Omega_n$ and $\rho^{lm}_n$ with the right number of derivatives. 
The same argument applies to the $\psi_n$.

  To find the master Legendre transform    
 which takes the infinite number of differential equations [9,21] for the $\Omega_n$ into an infinite
number of differential equations for the $\rho^{lm}_n$ is a very difficult problem. All we can
do is outline some of the most important feautures and propose to embed the $SU(2)$ solutions into the $SU(\infty)$ ones as a special simpler case.

Another  discrepancy  which at first  sight might be troublesome  
is that the number of variables of  $\Omega$  and $ \rho$ does not match.  The former has four
variables whereas the latter has five.  It is only after the $q,p$ dependence is factorized as shown in
(35)  that one can match the variables in $\Omega_n (r, z',{\tilde z}')$  with those in  $\rho^{lm}_n
(t, z_+, z_-)$.  However, the $\psi_n (z_+, z_-, t)$ belonging to the $SU(2)$ Moyal Toda equation  (29)
admit a perfect match with the    $\Omega_n (r, z',{\tilde z}')$ .  For this reason, when one speaks
of the master Legendre transform from the $\Omega$,  obeying  the rotational Killing symmetry reductions
of the Moyal heavenly equation, (38),   to the Moyal Toda equation   one must refer to Strachan's 
$SU(2)$ Moyal Toda equation (29) as discussed at the beginning of this subsection.  

Despite these problems, 
the expansion in spherical harmonics given by eq-(35) is  mathematically sound. Since derivatives of
spherical harmonics w.r.t. the angle variables can be reexpressed in terms of sums of spherical
harmonics,  plugging the value of $\rho$ given by the r.h.s  of  (35)  into eq-(32)  yields an
infinite number of  differential equations for the functions $\rho^{lm}_n (\tau,t)$ that in principle
could be solved by iterations.  A knowledge of  the solutions $\rho (\tau,t,q,p,\hbar) $ to eq-(32)  
will allow to compute the operator $ { \hat \rho} (\tau,t)$ ,  obeying the operator equations of motion
(31),   by recurring to the inverse symbol map which converts real functions in phase space into self
adjoint operators acting  in  the Hilbert space of square integrable functions on the line.  This is
precisely what is meant by a WWGM quantization of the continuous Toda field.  To solve (32) using (35) 
is clearly a very complicated matter.

An interesting question ,  which simplifies matters a little bit,  is whether or not  one can find a
particular subset of solutions to the $\rho^{lm}_n$  which solve the continuous Moyal Toda equation  
in such  a way that  the following factorization  condition occurs :

$$\sum_{l=0}^{l=n} \sum_{m=-l}^{m=+l} \rho^{lm}_n Y_{lm} =\psi_n f_n (q,p)=\rho_n (z_+, z_-,t,q,p).
\eqno (40)$$
( no sum over $n$ ) where the functions  $\psi_n $  are precisely those solving  Strachan's equation  (29)  after expanding
$\psi$ in powers of $\hbar$; i.e.  one has embedded the $SU(2)$ Moyal  Toda  into  the continuous Moyal
Toda .    Solving for $\rho^{lm}_n $ in (40) yields :

$$\rho^{lm}_n ( z_+, z_-, t) \sim  \psi_n (z_+, z_-, t) 
\int d\varphi d(cos\theta) Y_{lm}(\varphi, \theta) f_n (q,p) . \eqno (41)$$
where $f_n (q,p)$ are functions which in principle can  be determined  from (32) after inserting  eq-(40) into ( 32).  
This is clearly a very difficult task. 

To conclude this subsection,  to establish a master Legendre transform  from eq-(38) to eq-(32) 
runs into the discrepancy  in the number of derivatives w.r.t the $z_+, z_-$ or $\tau$ variable
appearing in the l.h.s of eq-(32) compared to the arbitrarily high number of derivatives w.r.t the
${\tilde z}'$ variable appearing in  eq-(38) [9,21].   After  counting
derivatives at any given order $n=N$,  acting on  the quantity $\Omega_N$,  that is  determined 
iteratively in terms of the previous $\Omega_0,.....\Omega_{N-1} $ ones,   reveals that this
discrepancy  does not seem to represent a major problem.  One  could  forsee  the possibility of 
modifying  the number of  derivatives    appearing in the l.h.s of eq-(32) that includes derivatives
of  arbitrary order in $\tau$ or $z_+, z_-$ . At the moment this seems unnecessary. 
The explicit master Legendre transform can be constructed in the very special case when one embeds the $SU(2)$ solutions ( of eq-(29) ) into  the $SU(\infty)$ case ( eq-(32)).   

Concluding, by sorting out the signature subtelties used by different authors [3,8], the transformation  
from $\Omega$, after a rotational Killing symmetry reduction,  to the $SU(2)$ Moyal Toda field $\psi$ is attained 
in three step-process by using eqs-(16,19,28). In this fashion one establishes the sought-after  
$\Omega=\Omega [X_i (\psi (\tau,t,\hbar),z(\tau,t,\hbar))]$ relation due to the fact that in this case 
the YM potentials $A_i=X_i$.  
This is probably one of the most relevant result of this work. 

\bigskip

\centerline {\bf IV. The Generalized Moyal Nahm Equations} 
\centerline {\bf Moyal Deformations of Loop Algebras, Symplectic Diffs in 
$4D$} 
\centerline {\bf And Continuum Lie Algebras .}

We have been studying $reductions$ of
the generalized  Moyal Nahm equations related to the Moyal deformations of the 
$SU(\infty)$ ASDYM/SDYM equations in $4D$, an effective $6D$ theory. The most
general Moyal Nahm equations require , at least, an extra set of $q',p'$
coordinates and hence one has an effective $8D$ theory where the Moyal bracket
is
taken w.r.t the enlarged phase space. i.e; the Weyl Wigner Moyal formalism
involves mapping operator valued quantities living in a $4D$ spacetime 
( belonging to a Hilbert space of  $L^2 (R^2)$, instead of
$L^2(R^1)$)   into functions of the enlarged phase
space, $q,p,q',p'$. Instead of dealing with Moyal deformed symplectic diffs of
a two-dim surface one is now dealing with Moyal deformed symplectic diffs in $4D$. Therefore the
effective theory is now $8D$ instead of $6D$ !

We $define$ the generalized  Moyal Nahm equations as :

$$\epsilon _{\alpha\beta\gamma}{ d{\cal T} _\gamma \over d \phi} =
\{\{ {\cal T}_\alpha, {\cal T}_\beta \}\}.~~~{\cal T}_\alpha [\phi
(x^\mu;q,p,q',p',\hbar);q,p,q',p';\hbar]. \eqno ( 42) $$
where now the Moyal bracket must be taken w.r.t an enlarged phase space
$q,p,q',p'$.  There is an explict  and implicit $q_i,p_i,\hbar$ dependence in the Moyal-Nahm fucntions. The Moyal bracket of $f,g$ w.r.t the enlarged phase space is  compactly written as : 

$$    \{\{ f,g\}\}\equiv {1\over \hbar} f (sin [\hbar (   {\overleftarrow \partial}_q {\overrightarrow
\partial}_p - q\leftrightarrow p + {\overleftarrow \partial}_{q'} {\overrightarrow \partial}_{p'}
-q'\leftrightarrow p')])g. \eqno (43)$$
expanding the sine  function in powers of $\hbar$ one retrieves the infinite derivative terms. 
The WWM map takes self-adjoint operator-valued quantities, living 
in the Hilbert space $L^2 (R^2)$, 
${\hat A} (x^\mu)$,  into real valued functions in phase space ${\cal A} (x^\mu; q,p,q',p';\hbar)$

$${\cal A} (x^\mu; q,p,q',p';\hbar)\equiv 
\int^\infty_{-\infty}  d^2\xi  <{\vec
q}-{{\vec \xi} \over 2}|{\hat A} (x^\mu)|{\vec q} +{{\vec \xi} \over 2}>
exp [{i {\vec \xi}.{\vec p}\over \hbar } ]. ~{\vec q}=(q,q').~{\vec p}
=(p,p').\eqno (44a)$$    

$$|{\vec q} +{{\vec \xi} \over 2}>=|{q}_1 +{{ \xi}_1 \over 2}>\otimes~ 
|{ q}_2+{{\xi}_2 \over 2}>...... \eqno (44b)$$
Imagine representing  now the $SU(2)$ Lie-algebra valued YM potentials 
( matrices) in terms of operators in ${\hat q}, {\hat p}, {\hat q} ', {\hat p}'$
and performing afterwards the WWM map ( (44a). 

 The relevant algebra is now the Moyal deformations of symplectic diffs in $4D$
instead of $2D$. For this reason it is incorrect to say that one has
"$SU(\infty)$" Moyal Nahm equations. For example, the generators are labelled as $V^{l,{\vec k}}_m$
where 
${\vec k}=(k_1,k_2) $ [20] and obey the infinite dimensional generalization 
of the $w_\infty$ algebra ( area-preserving diffs of the plane)  :

$$[ V^{l,{\vec k}}_m,  V^{j,{\vec l}}_n] = [(j+1)m-(l+1)n]
V^{l+j,{\vec k}+{\vec l}}_{m+n} +{\vec k} \times {\vec l}~ 
V^{l+j+1,{\vec k}+{\vec l} }_{m+n}. \eqno (45) $$
This algebra of symplectic diffs in $4D$ has
a realization in terms of ordinary Poisson brackets w.r.t the
$q,p,q',p'$ enlaged phase space variables.  The Moyal deformations are
obtained by replacing ordinary Poisson brackets by Moyal ones. If one had a
representation of the $SU(2)$ algebra as linear operators in $L^2 (R^2)$,  instead of the known
representations in $L^2 (R^1) $ [10],  one could then evaluate the WWM map 
(44a)
and obtain solutions to the generalized  Moyal Nahm equations 
as it was done in  [10].

Another type of generalized Moyal Nahm equations one could write  is such  where instead of
having a Moyal bracket w.r.t the enlarged phase space one has  a
$partial$ Moyal bracket w.r.t one set of $q,p$ variables :

$$\epsilon _{\alpha\beta\gamma}{ d{\cal T} _\gamma \over d \phi} =
\{ {\cal T}_\alpha, {\cal T}_\beta \}_{q,p}.~~~{\cal T}_\alpha [\phi
(x^\mu;q,p,q',p',\hbar);q,p,q',p';\hbar]. \eqno (46) $$
The $problem$  with eq-(39) as such is that it does $not$ determine the 
functional dependence
on the $q',p'$ variables since there is no differential operators which involve
now the $q',p'$ variables. For this reason we should $disregard$ (46) as a valid
equation.

We shall study the  reductions of  (42) with the goal of obtaining the continuous Moyal 
Toda equation  (32). Firstly ,  one chooses  $\phi =\tau$ then
the anstaz (7) gives $A_0=0, A_i\sim {\cal T}_i$ . Secondly, imposing the
reduction condition  $t=q'$ while recurring to an ansatz   that
allows one to decouple the $cosp', sin p'$ terms , 
after computing the Moyal
bracket in (42),  giving,  finally,  an
equation involving $only $ the $\tau,t,q,p$ variables . If we set :  

$${\cal T}_1=R(\tau,t=q',q,p,\hbar)cosp'. ~{\cal T}_2=R(\tau,t=q',q,p,\hbar)sin p'.
~{\cal T}_3=z(\tau,t=q',\hbar).\eqno (47)$$
After plugging (47) into (42), the terms in  $cosp',sinp'$ decouple and eliminating $z$  one obtains
the following highly nontrivial equation  for the function $R$,  after computing the Moyal bracket w.r.t
the enlarged phase space variables  :

$${\partial^2 lnR \over \partial \tau^2} =
\alpha ({\Delta -\Delta^{-1} \over \hbar})^2 R^2 +\beta ({\Delta -\Delta^{-1} \over \hbar})
\sum_{n=1}^\infty (\hbar) ^n C_{i_1...i_n}(\partial _{i_1} .......\partial_{i_{n-1} }R) (\partial
_{j_1}........\partial_{j_{n}} R). \eqno ( 48 )$$
The second terms of  the r.h.s of  (48) contain mixed derivatives of infinite order w.r.t  the $t,q,p$
variables.  $\alpha, \beta$ and $C_{i_1....i_n} $ are constants.  Strachan'  $SU(2)$ Moyal Nahm
equation is recovered  automatically by droppping the extra $q,p$ dependence  on $R$ (  so that the
second term in the r.h.s of  (48) becomes zero)   and by equating $R^2=e^\rho$.  The l.h.s  is then
equal to the l.h.s of  (29).

 Due to the extra $q,p$ dependence the sutuation changes drastically. 
In order to obtain  an equation like (32) it is required to establish the $new$ functional relation
between the $R$ and $\rho$ functions.  The previous  one $R^2 =e^\rho$ does not work .  The situation
now is more complex if one wishes  to perform the one-to-one WWM map taking the operator form of the
equations given by eq- (31) into  the Moyal Toda equation  (32).  
For example, if one sets : 

$$R*R =e^{*\rho}  \Rightarrow { \rho \over 2} \not= ln R. \eqno (49a) $$
the l.h.s of eq-(48) will no longer be : 

$${1\over 2} {\partial ^2 \rho \over \partial \tau^2}. \eqno (49b)$$
Furthermore,  if  (49a)  were the correct relation between $R$ and $\rho$,  the r.h.s of  ( 48 )  would 
$not$  equal  the r.h.s of  (32). Although the r.h.s of  (48) contains the same type of  infinite
derivative terms as eq-(32) does , after using the relationship (49a),    the coefficients  $differ$ .
Therefore, if  (49a)  were satisfied the r.h.s of  (48) would  not equal that of  ( 32 ).

Assuming that (49a ) were  the correct relation between $R$ and  $\rho$, another obstacle is that the l.h.s
of  (48 )  should have  been  of the form :

$${\partial ^2 ln_* R \over \partial \tau^2} ={1\over 2} {\partial ^2 \rho \over  \partial \tau^2}.
\eqno (50a)$$ Where the $star$ logarithm and $star$ square root are defined :

$$ ln_*(e^{*\rho})\equiv \rho.~~~R*R =e^{*\rho} \Rightarrow R=[e^{*\rho}]_*^{1/2}. \eqno (50b)$$

In this fashion  the inverse symbol map corresponding to (48) would have  had the required operator
form. It is not difficult to see that  both sides of  eq- (48) do not have the adequate form to match
eq-(32).  Therefore, the anstaz   proposed in  (47) does not  work if one wishes to  to find a direct
relation between $R$ and $\rho$ .  What is required is to introduce  two auxiliary functions discussed 
below.  
The source of the problem is due to the new functional relationship between $R$ and $\rho$ 
that will render both sides of eq-(32) correctly.   Nevertheless matters are not final. Eq-(48) is 
per se satisfactory in the sense that it yields a well defined differential equation for the $R$
function with an infinite number of derivative terms.  Imposing  the condition :

$${1\over 2} {\partial ^2 \rho \over \partial \tau^2}  = 
{\partial^2 ln R \over \partial \tau^2}           \eqno (51)$$
implies that  $\rho/2$ and $ln R$ now differ by : 

$$ {1\over 2} \rho = lnR + F(t,q,p,\hbar) \tau + G(t,q,p,\hbar). \eqno (52)$$
This fixes $\rho$ in terms of $R$ and two auxiliary functions, $F,G$ .  Solving for $R$ yields :

$$R =e^{\rho/2 -F\tau -G}=e^{\rho/2} e^{-F(t,q,p,\hbar)\tau-G(t,q,p,\hbar)}. \eqno (53)$$
Inserting this  new value for $R$ into  the r.h.s of  (48) and equating it to the r.h.s of  (32)
determines another  differential equation for the ${ F,G} $ functions   in conjuction with the original
Moyal Toda equation.  Three  coupled differential equations for $\rho$ and ${F,G}$ are obtained in this
fashion given by  the three eqs- (32,48,51). It would be erroneous to equate the r.h.s of (48) to the r.h.s of (26) because the number of derivatives w.r.t the $t$ variable does $not$ match. In (48) there is an infinite number whereas in (26) they are only quadratic.  

Therefore,    one can in principle obtain  the  continuous Moyal Toda  equation from axial 
symmetry reductions of the Moyal Nahm equation (42,47)  if, and only if, one introduces  two auxiliary
functions, $F,G$  that are determined, in conjunction with $\rho$, by  three  coupled differential
equations  (32,48,51).  Whether this system of three coupled differential equations   is  
compatible and consistent   is another matter.  This very difficult question remains to be
answered in addition to the uniqueness and existence of solutions.  This  implies  that for every
solution
$\rho$ to eq-(32) the remaining  two differential equation (48,51) yield  a well defined nontrivial
${F}(t,q,p,\hbar)$ in terms of $
\rho$ .  Similar arguments applied to the remaining auxiliary function $G(t,q,p,\hbar)$.

It is also reasonable to look for other ways of embedding the Moyal-Toda equations into 
the Moyal-Nahm equation that do $not$ require a direct decoupling of the $cosp',sinp'$ functions, for
example.  Inotherwords,  integrating out the $p'$ variable without the need to reduce the generalized
Moyal-Nahm equations.  
A plausible  embedding ( it is problematic)   could have been as follows :

$$ \int_{-\infty}^{+\infty}  dq'\int_{-\infty} ^{+\infty} dp' {\partial {\cal T}_3 \over
\partial \phi}={\partial ^2
\rho (\tau,t,q,p,\hbar) \over \partial \tau^2}. \eqno (54a)$$
$$ \int _{-\infty}^{+\infty} dq' \int_{-\infty} ^{+\infty} dp'~\{\{ {\cal T}_+, {\cal T}_-\}\}={1\over
4}  ({\Delta -\Delta^{-1} \over \hbar})^2 e^{*\rho} =$$  
$$ { \partial \over \partial t^2}   e^{*\rho (\tau,t,q,p,\hbar)}+{1\over 3} (\hbar)^2 
{\partial \over \partial t^4} e^{*\rho (\tau,t,q,p,\hbar)} +......\eqno (54b)$$   
with ${\cal T}_{\pm} ={\cal T}_1 \pm i {\cal T}_2 $  and  $\phi =f(\tau + it)+g(\tau -it)$  contains 
implicitly the
$\tau,t$ dependence of $\rho$.  The three functions ${\cal T} _\alpha$ are required to obey the 
generalized Moyal-Nahm equations  (42). 
One has  some  vanishing boundary terms due to total derivative terms but $not$ all of the terms are 
total derivatives due to the integration w.r..t half of the phase space  variables.  However there is a
major $problem$ with the above embedding :  there are derivatives of infinite order w.r.t the $t$
variable in the r.h.s of eq-(54b) whereas  the derivatives w.r.t the function $\phi$ are of  first order
only.  Since $\partial _t=(\partial \phi/\partial t) {\partial \over \partial \phi}$ the above
equations are inconsistent.   There is no place where derivatives of infinite order w.r.t the function
$\phi$ appear . Therefore we must disregard the above embedding.

The use of the ${\cal T}_\alpha$ was
sucessful in rendering the  ordinary $SU(2)$ Moyal Toda equation ( 29 )  from the Moyal Nahm
equations as discussed earlier when one imposed the reduction $q=q',p=p'$ conditions and set $\phi=\tau,
t=q$. Nevertheless, despite the above problems, the putative continuous Moyal Toda equation (26) admits the following embedding into the
reductions of the Moyal Nahm equations ,  after setting for example  $\phi =f+g=\tau
\pm it$ ,     :

$$ {\partial {\cal T}_3 \over \partial \phi}={\partial^2 \rho  \over \partial \phi^2} = \rightarrow 
\rho (\phi,q,p,\hbar)  =\int {\cal T}_3 d\phi  +F(q,p,\hbar)\phi+G(q,p,\hbar). \eqno (55a)$$ 

$$\{ {\cal T}_+, {\cal T}_- \} = -{\partial^2 \over \partial \phi^2} e^{* \rho (\phi,q,p,\hbar)}.~~~
\rho=\rho (\phi,q,p,\hbar).~~~q=q', p=p'. \eqno (55b)$$
Following similar arguments as above yields  a system of three diferential equations, eqs-(26,55a,55b)
to solve for the three functions $\rho (\phi, q,p,\hbar) ,F,G$.

 Finally, the embbeding of the  continuous Moyal Toda equation (32)  into the  generalized Moyal Nahm
requations (42)  which is more directly linked to eq- (47) is to set $q'=t$ and  to fix the function
$\phi=\tau $ so that the ansatz in (7) yields $A_0 =0$ and $A_i (\tau,q'=t,p',q,p,\hbar)={\cal T}_i$
for $i=1,2,3$.
Given a solution to the generalized  $SU(2)$ Moyal Nahm equations (using Moyal brackets w.r.t the extended phase space variables)   for the three potentials 
${\cal A}_i$, one selects the particular equation    :

$${\partial {\cal A}_3 \over \partial \tau } = \{\{ {\cal A}_+, {\cal A}_- \}\}.~~~
{\cal A}_{\pm} ={1\over \sqrt 2} ({\cal A}_1 \pm i {\cal A}_2) . \eqno (56a) $$ 

A  partial integration taken w.r.t the $p'$ variable only  yields the embedding relations :

$$ \int_{-\infty}^{+\infty}  dp' {\partial {\cal A}_3 \over \partial \tau }={\partial ^2 \rho
(\tau,t,q,p,\hbar) \over \partial \tau^2}. \eqno (56b)$$

$$ \int_{-\infty}^{+\infty}  dp'~\{\{ {\cal A}_+, {\cal A}_-\}\}=
{1\over  4}  ({\Delta -\Delta^{-1} \over \hbar})^2 e^{*\rho}= 
\partial_t^2 e^{*\rho (\tau,t,q,p,\hbar)}+ {1\over 3} (\hbar)^2 \partial_t^4
e^{*\rho} +............    \eqno (56c)$$
Both sides of eq-(56c)  now contain derivatives of infinite order w.r.t the $q,p,q'=t$ variables and no
inconsistency arises.   There are some total derivative terms  w.r.t $p'$  which vanish  after
integration but not all of the terms appearing in the l.h.s of  (56c) are total derivatives.   
Eqs-(56) are  consistent in their structure.  
From eq-(56b) one learns that :

$$\rho =\int_{-\infty}^{+\infty} dp'\int^ {\tau'=\tau}_{\tau'=0}{\cal A}_3 d\tau '  +{\cal F}
(t,q,p,\hbar)\tau+ {\cal G} (t,q,p,\hbar). \eqno (57)$$
The functions ${\cal F} ,{\cal G} $ are  not arbitray but are  part of the system of 
three coupled differential equations  given by  eqs-(32,56b,56c) ; i.e  ${\cal F}, {\cal G}$ are given
in terms of the ${\cal A}_i$ satisfying  the
generalized Moyal Nahm equations (42). 

For instance,  inserting (57) into  (32)  and (56c) determines the differential equations  
for ${\cal F} (t,q,p,\hbar)$  and $ {\cal G} (t,q,p,\hbar)$ in terms of  the  fields ${\cal A}_ i   (
\tau,q'=t,p',q,p,\hbar)$  which are solutions to  eq-(42) .  The embedding is characterized by
integrating out the variable $p'$ without imposing a reduction on the eqs- (42)  which would decouple 
the $p'$ variable directly .  In all these embeddings , two auxiliary  functions are required.

 It is warranted to see whether it is possible to find a 
Killing symmetry reduction of the generalized Moyal Nahm equations (42) directly to the Moyal Toda
equations  without the need to recur to auxiliary functions and  avoid the complicated set of 
 three coupled differential
equations (32,56b,56c).  The essence of the problem lies in the fact that the $t$ variable plays two
different roles.  In one case,  like in the rotational  Killing symmetry reduction of the heavenly
equation,  it  behaves like an ordinary spacetime variable and in another  it behaves like an 
$internal$  phase space variable associated with the Lie algebra of  the area-preserving diffs of the
sphere.  It seems very difficult to reconcile both roles within  the WWGM formalism without recurring
to the coupled system of differntial equations.

Finally, we shall discuss the ansatz  that suceeds  in 
rendering reductions of the generalized Moyal-Nahm equations in the form prescribed by eq- (32)
without the introduction of auxiliary functions.  We believe that there maybe a reduction  ( other than
the axial symmetry reduction proposed above )  of the generalized Moyal-Nahm equations, (42), 
that sucessfully decouples the $p'$ variable and that reproduces eq-(32) without the auxiliary
functions.   

Essentially one has the data $\{ {\cal T}_\alpha \}$ for three Moyal-Nahm functions functions and three functions $\{{\cal L}, {\cal H}, {\cal M}\}$ required in the Lax-Brockett formalism of continuum {\bf Z} graded Lie algebras [13]. Eq-(19) establishes the correspondence among these data. However, this is only possible if one can construct the three twistor-like 
transformations  which map the now-deformed scalar field , $\phi (x^\mu; q_i,p_i,\hbar)$ into the now-deformed three functions appearing in eqs-(2,3,4) : $\rho, u, \kappa$ depending on $\tau,t;q_i,p_i,\hbar$. One of them is the Moyal deformed continuous Toda field : $\rho$.  
Presummably this should be related to the problem of deformations of twistor surfaces and Kodaira-Spencer deformation theory [16]. Unfortunately, we cannot say more on this matter. A start will be in constructing Moyal deformations of continuum Lie algebras. We are unaware if this has ever been done.

Roughly speaking, an example of the analog of the twistor-like transformation is to introduce two $functionals$ , ${\cal F}_+, {\cal F} _3$ so that the correspondence in (19) implies:

$$  {\cal F}_+[ {\cal T}_1 (\phi (\tau, t;q_i,p_i\hbar);q_i,p_i,\hbar )) + i{\cal T}_2 ((\phi (\tau, t; q_i,p_i\hbar);q_i,p_i,\hbar )) ]={\cal L}[u,\rho; q_i,p_i,\hbar].\eqno (58)$$
. Lets assume for simplicity that the functional in (58) is linear , ${\cal F}_+ [A+B]= {\cal F}_+ [A]+ {\cal F}_+ [B]$. The other expression is :  

$$  {\cal F}_3[i{\cal T}_3 (\phi (\tau, t; q_i,p_i\hbar);q_i,p_i,\hbar ))    ]=
{\cal M}[\kappa ; q_i,p_i,\hbar].\eqno (59)$$
with $u,\rho,\kappa$ depending on $\tau,t,q_i,p_i,\hbar$ and the deformed $\phi$ obeys a deformed Laplace equation obtained by the Moyal quantization of the Ivanova-Popov construction. If, and only if, the representations of the {\bf Z}-graded continuum Lie algebras were known, then one would have a knowledge of the explict functional relation of ${\cal L}, {\cal M}$ in terms of $u,\rho,\kappa$ that can be obtained from eqs-(2-6) after performing the WWGM map. Similarily, the three functions ${\cal T}_\alpha$ are solutions to the generalized Moyal-Nahm equations (42). Therefore, eqs-(58,59) would have  been  the defining expression for the two functionals, ${\cal F}_+, {\cal F}_3$. Unfortunately we do $not$ know the explicit form of ${\cal L}, {\cal M}$ because we lack the knowldedge of the representations. All we can do to find out the expressions for     ${\cal F}_+, {\cal F}_3$     is the following :

Upon establishing the correspondence between the Moyal-Nahm eqs-(42) and the Lax-Brockett equations (18,19) ( with the crucial difference that now one 
must use the Moyal bracket w.r.t the enlarged phase space coordinates) gives after the chain rule and due to the fact that we have assumed for simplicity that the functional ${\cal F}_+$ was $linear$ in its arguments :

$${\partial {\cal L} \over \partial \tau} = 
({ \partial \phi \over \partial \tau})
  [   {\partial  {\cal T}_1 \over \partial \phi} 
{\delta {\cal F}_+ \over \delta {\cal T}_1 } +i {\partial  {\cal T}_2 \over \partial \phi}{\delta {\cal F}_+ \over \delta {\cal T}_2 }    ]=$$  
$$({ \partial \phi \over \partial \tau}) [    {\delta {\cal F}_+ \over 
\delta {\cal T}_1  } \{\{ {\cal T}_2,  {\cal T}_3  \}\} + 
 i {\delta {\cal F}_+ \over 
\delta {\cal T}_2  } \{\{ {\cal T}_3,  {\cal T}_1  \}\}    ] =
\{\{ {\cal L}, {\cal M}  \}\}.\eqno (60)$$

If one chooses the identity functionals for both ${\cal F}_+, {\cal F}_3$ then as expected one arrives at the constraint in eq-(20). The correspondence turns into an strict identification. Avoiding this trivial and restrictive special case, we see then that in general the $ {\cal L}, {\cal M} $ expressions in terms of $\phi$ are given by the functional map in eqs-(58,59). Hence, the  last eqs-(58-60) are the defining highly nontrivial differential equations for the two functionals ${\cal F}_+, {\cal F}_3$, once the defining equation for the deformed $\phi$ is known and solutions of the generalized Moyal-Nahm equations (42) for the three Moyal-Nahm functions , ${\cal T}_\alpha$ have been found. Clearly, this is a extremely complicated system of equations to solve. Instead of having two auxiliary functions we have in this case two functionals. For this reason, it is of tantamount importance to construct representations   
of continuum Lie algebras and for that matter representations of $SU(\infty)$ as well.

To sum up :  we have shown that axial symmety reductions of the generalized Moyal Nahm equations 
(47)  in principle yield the continuous Moyal Toda equation  (32) with the provision  that  a coupled
system of  three differential equations  for $\rho, {F,G} $ is  solved, eqs-(32,48,51).   An
embedding   of (32) into (42)  
is also possible after integrating out one of  the phase space variables. One is also required to
solve  three coupled system of complicated differential equations, eqs-(32,56b,56c),  for $\rho$ and the two 
new auxiliary functions ${\cal F}, {\cal G}$  .  
Eqs-(58-60) depict another way of obtaining Moyal Toda equations directly from the Moyal Lax formalism (19) and the Moyal Nahm equations (42). Again, two functionals must be introduced which are determined by a very complicated system of differential equations.

Other Moyal deformations are those related to the 
infinite dimensional loop algebras
associated with $w_\infty$ algebras.  For example, the loop algebra of
$sdiff(R^2)$ , the algebra of maps of the circle into $w_{\infty}$ , in the basis of functions $x^{s+m}
y^{s-m} $ is :

$$[v^s_m (\sigma), v^t_n (\sigma ') ] =[(t-n)(s+m)-(s-m)(t+n)]v^{s+t-1}_{m+n}
(\sigma) {\partial \over \partial \sigma} \delta (\sigma -\sigma'). \eqno (61)
$$ 
these loop algebras may admit Moyal deformations as well since the Moyal deformation of the
centerless $w_\infty$ algebra $is$ the centerless  $W_\infty$ algebra
[17,18,19,20]. Central extensions
can be added as well [19]. Hence,  the Moyal deformations of the algebra (42)
will be just the infinite dimensional loop algebra associated with $W_\infty$.

Finally, Moyal deformations of the {\bf Z} graded continuum Lie algebras [13] ought to be very
relevant in the Moyal quantization program of the continuous Toda theory.
Especially in regards to determining the differential equations for the $u
(\tau,t,q,p,\hbar)$ and $\kappa (\tau,t,q,p,\hbar)$ appearing in 
(4,18,19) and in obtaining eq-(32) directly from the Lax-Brockett double 
commutator formalism. More on this
shall be said in a forthcoming publication.  

\bigskip

\centerline {\bf V.  Conclusion} 

\smallskip

We have explictly presented a class of solutions to the Moyal $SU(\infty)$ ASDYM
equations in four dimensions  that are related to the $reductions$ of the
generalized Moyal Nahm quations via the Ivanova-Popov ansatz.  A dimensional 
reduction yields solutions to the Moyal deformations of the ASDG equations. The SDYM and  SDG case
requires a separate study.  

Since the ASDYM equations studied by Ivanova and Popov [3] in Euclidean $4D$ correspond to the SDYM equations in $2+2$ dimensions studied by [8], 
one can  write down the master
Legendre transform that maps the rotational Killing symmetry reductions of the Moyal heavenly
equations given by eq-(14) into the $SU(2)$ Moyal Toda equations given by eq-(29). A three step process is required to attained such a map and it is explicitly given by eqs-(16,19,28). This is one of the most relevant results of this work. 
Solutions to the continuous Moyal Toda equation may be obtained by embedding the $SU(2)$ solutions into the $SU(\infty)$ case. In this fashion, a Legendre map from $\Omega$ to the continuous Toda field $\rho$ can be attained.

Three different types of Toda equations have been studied. The   $SU(2)$  and continuous  Moyal Toda
equations, eqs-(26,29,32) ,   have been explicitly  derived .
To a first order approximation, neglecting the deformations of the scalar $\phi$,  the  $SU(2)$ Moya Toda equations  can be simply obtained from the $SU(2)$ Moyal Nahm equations  
when  ( the undeformed scalar)  $\phi =f+g=\tau$ and $t=q$.  
Finally,  the 
generalized   Moyal Nahm equations  (42)  have been provided that 
$contain$ the continuous Moyal Toda equations  (32)  after a
suitable reduction,  similar to the one 
performed by  Strachan [16]  which yields the $SU(2)$ Moyal Toda from the $SU(2)$ Moyal
Nahm equations. This reduction requires the introduction of two auxiliary fields.  Further details of
this  reduction  is currently under investigation.  
Embeddings of
the various forms of the Moyal Toda equations  into the  Moyal Nahm equation 
were  also provided  and,  again, the introduction of two auxiliary fields was required.

The proyect for the future is to study eqs-(16,19) and the generalized Moyal-Nahm equations (42) as shown in eqs-(58-60). But now one should use the appropriate $deformed$ scalar field $\phi (x^\mu;q_i,p_i,\hbar) $ 
satisfying deformations of Laplace equation. This will provide the indirect map between $\Omega$ and $\rho$ via the Moyal-Nahm-Lax pair formalism. Unfortunately we lack an explicit knowledge of the form of the ${\cal L}, {\cal M}$ expressions ( given by eq-(19)) in terms of the deformed $\rho,u,\kappa$ fields. If we did, then one could establish the sought-after correspondence after  using eq-(19) and eqs-(16). 

The main obstacle is the construction of the defining equation for the full deformed scalar $\phi (x^\mu;q_i,p_i,\hbar)$ that must be obtained by an explicit Moyal quantization program of the Ivanova-Popov construction. In the $\hbar =0$ limit the $\phi (x^\mu,q_i,p_i,\hbar)\rightarrow \phi (x^\mu)$ obeying the original (undeformed) Laplace equation.   
Presumably, this could be a realization of deformations of twistor surfaces 
. The connection to  Kodaira-Spencer deformation theory [16] is unknown at the moment.

Other Moyal deformations applied to higher extended objects, $p$-branes , remain
to be studied : the so called Moyal-Nambu-Poisson Algebras related to
deformations of the volume forms. The natural
deformation quantization technique is the Zariski product [22] 
which generalizes the
Moyal product to $p$-branes. Octonionic [30]  and Quaternionic Moyal Nahm equations
can be constructed as well using the octonionic/quaternionic structure constants
instead of the $\epsilon_{\alpha\beta\gamma}$ tensor density. 
The fact that the generalized  Moyal Nahm equations require $8D$ may have an
important role in undertanding the quantum dynamics of the $11D$ membrane [23]
and the role of $W_\infty$ algebras [21] .

\centerline{\bf Acknowledgments}

We are indebted to M. Przanowski for valuable discussions and to T. Ivanova for useful remarks.   
To J. Boedo and L. Russo for their help and hospitality at the University of California, San Diego. 
This work was supported in part by the CONACyT, Mexico.

\centerline{ \bf References}

1-C. Castro : Phys. Lett {\bf B 288} (1992) 291.

Journ. Chaos, Solitons, Fractals, {\bf 7} no.7 (1996) 711.

``The Noncritical $W_\infty$ String Sector  of the Membrane `` hep-th/9612160.

2-. J.Hoppe :``Quantum Theory of a Relativistic

 Surface `` M.I.T Ph.D thesis  (1982).

3-T. A. Ivanova, A. D. Popov : Jour. Math. Phys {\bf 34} (2) (1993) 674.

T. A. Ivanova, A. D. Popov : Theor. Math. Phys {\bf 94} (2) (1993) 225.

Lett. Math. Phys {\bf 23} (1991) 29. JETP Lett {\bf 54} (1991) 69.

4- E.G. Floratos, G.K. Leontaris : Phys. Lett. {\bf B 223} (1989) 153.

5- G. Chapline, K. Yamagishi : Class. Quan. Grav. {\bf 8} (1991) 427.

 G. Chapline, K. Yamagishi : Phys. Rev. Lett {\bf 66} (23) (1991) 3064.

G. Chapline : Mod. Phys. Lett {\bf A 7} (22) (1992) 1959.

6-. E. Bergshoeff, E. Sezgin, Y. Tanni, P.K. Townsend : Ann. Phys. {\bf 199} (1990) 

340.

B. de Wit, J. Hoppe, H. Nicolai : Nucl. Phys. {\bf B 305} (1988) 545.

B.Biran, E.G. Floratos, G.K. Saviddy : Phys. Lett {\bf B 198} (1987) 32.

T. Banks, W. Fischler, S. Shenker and L. Susskind : Phys. Rev {\bf D 55} 

(1997) 5112. 

7- C. Castro : J. Math. Phys. {\bf 34} (1993) 681.

Jour. Math. Phys {\bf 35} no.6 (1994) 3013. 

8-J. Plebanski, M. Przanowski : Phys. Lett {\bf A 219} (1996) 249.

9- C. Castro ,  Phys. Lett.  {\bf 413 B} (1997) 53 . 

10- H. Garcia-Compean, J. Plebanski : `` On the Weyl-Wigner-Moyal description 

of $SU(\infty)$ Nahm equations'' hep-th/9612221.

11-H. Weyl : Z. Phys. {\bf 46} (1927) 1. E. Wigner : Phys. Rev. {\bf 40} (1932)

749.

J. Moyal : Proc. Cam. Phil. Soc. {\bf 45} (1945) 99. 

H. J Groenwold, Physica {\bf 12} (1946) 405. 

12- . V. Hussain : Jour. Math. Phys. {\bf 36}(12) (1995)6897.

T. Ivanova : " On Current Algebra Symmetries of the SDYM Equations"

hep-th/9702144. 

E. Alfinito, G. Soliani and L. Solombrino : " The symmetry structure of the

heavenly equations " hep-th/9604085.

A.D Popov, C.R Preitschopf : Phys. Lett {\bf B 374} (1996) 71.

A.D Popov, M. Bordermann, H. Romer : Phys. Lett {\bf B 385} (1996) 63.

13-M. Saveliev : Theor. Math. Phys. {\bf 92} (1992) 457.

M. Saveliev, A. Vershik : Phys. Lett {\bf A 143} (1990) 121. 

14- B. Fedosov : Jour. Diff. Geometry {\bf 40} (1994) 213.

M. Reuter: ``Non-Commutative Geometry of Quantum Phase Space`` hep-th/9510011

H. Garcia-Compean, J. Plebanski, M. Przanowski : `` Geometry associated with 

Self-Dual Yang-Mills and the Chiral Model approaches to Self Dual Gravity ``.

hep-th/9702046. 

5-C.D. Boyer, J.Finley III : Jour. Math. Phys. {\bf 23} (6) (1982) 1126. 

Q.H. Park : Int. Jour. Mod. Physics {\bf A 7} (1991) 1415.

16-I.A.B  Strachan : Phys. Lett {\bf B 282} (1992) 63.  Jour. Geom. Phys. {\bf

21} (1997) 255.  Jour. Phys. A {\bf 29} (1996) 6117.  Private communication.  

17- D.B. Fairlie, J. Nuyts : Comm. Math. Phys. {\bf 134} (1990) 413.

18- I. Bakas, B. Khesin, E. Kiritsis : Comm. Math. Phys. {\bf 151} (1993) 233.

19-C. Pope, L. Romans, X. Shen : Phys. Lett. {\bf B 236} (1990) 173.

20- I. Bakas, E. Kiritsis , in " Common trends in mathematics and quantum field

theories" . Eds. T.Eguchi, T. Inami and T.Miwa ( Prog. Theor. phys. Supplement ,

1990).

E.Sezgin : " Area-preserving Diffs , $w_\infty$ algebras and $w_\infty$ gravity"

hep-th/9202086. 

21-C. Castro : Phys. Lett. {\bf B 353} (1995) 201.

22- G. Dito, M. Flato, D. Sternheimer and L. Takhtajan : " Deformation

Quantization and Nambu Mechanics " hep-th/9602016.

23- D. Fairlie : `` Moyal Brackets in M theory        `` hep-th/9707190.

D. Fairlie, C. Zachos and T. Curtright : " Matrix Membranes and

Integrability", hep-th/9709042. 

24- C. Castro : in preparation. 

25- M. K Prasad , Phys. Lett {\bf 83B} (1979) 310. 

26-  G. W Gibbons , S. W Hawking : Phys. Lett {\bf 78B} (1978) 430.

T. Eguchi, A.J Hanson ,  Phys. Lett {\bf 74B} (1978) 249.

27-. M. Przanowski, S. Formanski : `` Searching for a universal integrable system ``

University of Lodz preprint. May, 1997.

28-. A. Dimakis and F. Muller-Hoissen, `` Integrable discretizations of chiral models via 

deformation of the differential calculus `` hep-th/9512007.  

``Deformations of classical geometries and integrable systems `` physics/9712002.

29-K. Takasaki, P. Guha, `` Dispersionless Hierarchies, Hamilton Jacobi Theory 

and Twistor Correspondences `` solv-int/9705013.

30- D, Fairlie , T. Ueno , `` Higher dimensional generalizations of the Euler Top 

Equations `` hep-th/9710079.

T. Ueno, `` General solution of the $7D$ octonionic top equations'' hep-th/9801079.

\bye

\centerline {\bf II The Moyal Quantization of the Continuous Toda Theory}

In this section we shall present the Moyal quantization of the continuous Toda theory. The Moyal deformations of the rotational Killing symmetry reduction of Plebanski self dual gravity equations in $4D$ were given by the author in [15] based on the results of [16]. Starting with  :

$$\Omega (y,{\tilde y},z,{\tilde z};\kappa )\equiv 

\sum_{n=0}^\infty (\kappa/{\tilde y})^n \Omega_n (r,z,{\tilde z}). \eqno (1)$$

where each $\Omega_n$ is only a function of the complexified variables $r\equiv y{\tilde y}$ and $z,{\tilde z}$ . Our notation is the same  from [15]. A real slice may be taken by setting ${\tilde y} ={\bar y}, {\tilde z}={\bar z},..$.   

The Moyal deformations of Plebanski's equation read  :

$$\{\Omega_z, \Omega_{{y}} \}_{Moyal} =1. \eqno (2)$$

where the Moyal bracket is taken w.r.t the  ${\tilde z}, {\tilde y}$ 

variables. In general, the Moyal  bracket may  defined as a power expansion in the deformation parameter, $\kappa$ :

$$\{f,g\}_{{\tilde y},{\tilde z}} \equiv [\kappa^{-1} sin~\kappa (\partial_{{\tilde y}_f}

\partial_{{\tilde z}_g} -  \partial_{{\tilde y}_g}

\partial_{{\tilde z}_f})]fg. \eqno (3)$$

with the subscripts under ${\tilde y}, {\tilde z}$ denote derivatives acting only on $f$ or on $g$ accordingly.

We begin by writing down the derivatives w.r.t the $y,{\tilde y}$ variables 

when these are acting on $\Omega$

$$\partial_{y}={1\over y}r\partial_r.~        

\partial_{{\tilde y}}={1\over {\tilde y}}r\partial_r.   \eqno (4)$$

$$\partial_y \partial _{{\tilde y}}=r\partial^2_r +\partial_r.~

\partial^2_{{\tilde y}}=({1\over {\tilde y}})^2 (r^2\partial^2_r +r\partial r).

$$

$$\partial^3_{{\tilde y}}=({1\over {\tilde y}})^3 (

r^3\partial^3_r +r^2\partial^2_r -r\partial r).

$$

$$...................................$$

Hence, the Moyal bracket (2)

yields the infinite number of equations after matching, order by order in $n$, powers of $(\kappa/ {\tilde y})$:

$$\{\Omega_{0z}, \Omega_{0{y}} \}_{Poisson} =1\Rightarrow (r\Omega_{0r})_r

\Omega_{0z{\tilde z}}-r\Omega_{0rz}\Omega_{0r{\tilde z}}=1. \eqno (5)$$

$$\Omega_{0z{\tilde z}}[-\Omega_{1r}+(r\Omega_{1r})_r]-

r\Omega_{1r{\tilde z}}

\Omega_{0rz}+\Omega_{1z{\tilde z}}(r\Omega_{0r})_r 

+\Omega_{0r{\tilde z}}(\Omega_{1z}-r\Omega_{1rz})=0. $$

$$......................$$

$$

.......................$$

the subscripts represent partial derivatives of the functions 

$\Omega_n (r=y{\tilde y},z,{\tilde z})$ for $ n=0,1,2.....$ w.r.t the variables   $r,z,{\tilde z}$ in accordance with the Killing symmetry reduction conditions. The first equation, after a nontrivial change of variables, can be recast as the $sl(\infty)$ continual Toda equation as demonstrated [2,3] . The remaining equations are the Moyal 

deformations. The symmetry algebra of these equations is the Moyal deformation of the classical $w_\infty$ algebra which turns out to be precisely the centerless $W_\infty$ algebra as shown by [19]. Central extensions can be added using the cocycle formula in terms of logarithms of derivative operators [20] giving the $W_\infty$ algebra first built by [21]. .

From now on in order not to be confused with the notation of [5] we shall denote for ${\tilde \Omega}(y',{\tilde y}',z',{\tilde z}';\kappa)$ to be the solutions to eq-(2). The authors [5] used $\Omega (z+{\tilde y},{\tilde z}-y,q,p;\hbar)$ as solutions to the Moyal deformations of Plebanski equation. The dictionary from the results of [15], given by eqs-(1-5), to the ones  used by the authors of [5] is 

obtained  from the relation :

$$\{ {\tilde \Omega}_{z'}, {\tilde \Omega}_{y'}\}_{{\tilde z}', {\tilde y}'}

 =\{ \Omega_w, \Omega_{{\tilde w}}\}_{{q,p}}=1.~\kappa=\hbar.~w=z+{\tilde y}.~

{\tilde w}={\tilde z}-y. \eqno (6a)$$

For example,  the four conditions : ${\tilde \Omega}_{z'}=\Omega_w;

  {\tilde \Omega}_{y'}=\Omega_{{\tilde w}}$ and ${\tilde z}'=q;{\tilde y}'=p$

are one of many which satisfy the previous dictionary relation (6a). One could perform a deformed-canonical transformation from ${\tilde z}',{\tilde y}'$ to the new variables $q,p$ iff the Moyal bracket $\{q,p\}=1$. Clearly, the simplest canonical transformation is the one chosen above. 

The latter four conditions yield the transformation rules from 

${\tilde \Omega}$ 

to  $\Omega$. The change of coordinates :

$${\tilde z}'=q.~{\tilde y}'=p.~z'=z'(w,{\tilde w},q,p|\Omega).~y'=y'(w,{\tilde w},

q,p|\Omega).\eqno (6b)$$

leads to :

$$z'=w+f(p,q).~y'={\tilde w}+g(p,q).$$

once one sets :

$${\tilde \Omega}[z'(w,{\tilde w}....);y'(w,{\tilde w}...,);{\tilde z}'=q;{\tilde y}'=p]=\Omega (w,{\tilde w},q,p). \eqno (6c)$$

for ${\tilde \Omega},\Omega$ obeying eqs-(2,6a). The $implicitly$ defined change of coordinates by the four conditions stated above is clearly dependent on the family of solutions to eqs-(2,6a). It is highly nontrivial. The reason this is required is because the choice of variables must be consistent with those of [9] to implement the WWM formalism.  

For example, choosing $\Omega=\Omega_o =z'{\tilde z}'+y'{\tilde y}'$ as a solution to the eqs-(2,5) yields for (6b) :

$$z'=w+{\lambda \over q}.~y'={\tilde w} -{\lambda \over p}. \eqno (6d)$$

The reality conditions on $w,{\tilde w}$ may be chosen to be :

${\tilde w}={\bar w}$ which implies ${\tilde z} ={\bar z};{\tilde y}=-{\bar y}$. It differs from the reality condition chosen for the original variables. It is important to remark as well that the variables $p,q$ are also complexified and the  area-preserving algebra is also : the algebra is $su^*(\infty)$ [4].

Now we can make contact with the results of [5,9]. 

In general, the expressions  that relate  the $6D$ scalar field 

$\Theta (z,{\tilde z},y,{\tilde y},q,p;\hbar)$ to the 

$4D~SU(\infty)$ YM potentials become, as a result of the dimensional reduction of the effective $6D$ theory to the $4D$ SDG one,  the following  [4,5] :

$$\partial_z \Theta =\partial_{{\tilde y}}\Theta =\partial_{{w}}\Theta.         ~~~\partial_y \Theta =-\partial_{{\tilde z}}\Theta=

-\partial_{{\tilde w}} \Theta. \eqno (7a)$$

with $\kappa \equiv \hbar$ and $w=z+{\tilde y};{\tilde w}={\tilde z} -y$. Eqs-(7a) are basically equivalent to the integrated dimensional reduction condition :

$$\Theta (z,{\tilde z},y,{\tilde y},q,p;\hbar)=\Omega (z+{\tilde y},{\tilde z}-y,q,p;\hbar )\equiv \Omega (w,{\tilde w},q,p;\hbar)\equiv \sum_{n=0}^\infty (\hbar)^n \Omega_n (r=w{\tilde w};q,p). \eqno(7b)$$

which furnishes the Moyal-deformed YM potentials : 

$$A_{{\tilde z}} ({\tilde y},w,{\tilde w} ,q,{p};\hbar)=

\partial_{{\tilde w}} 

\Omega (w,{\tilde w},q,p;\hbar) +{1\over 2}{\tilde y}.~ 

A_{{\tilde y}} ({\tilde z},w, {\tilde w} ,q,{p};\hbar)  =\partial_w \Omega (w,\tilde w,q,p;\hbar ) -{1\over 2}{\tilde z}.\eqno (8)$$. 

One defines the linear combination of the YM potentials  :

 $$A_{{\tilde z}}-A_{y}=A_{{\tilde w}}.

~A_{{\tilde y}}+A_{z}=A_{w} \eqno (9)$$

The new fields are denoted by $A_{w}, A_{{\tilde w}}$. After the following 

gauge conditions are chosen  

$A_z=0, A_y=0,$ [5] ,  it follows that $A_{{\tilde z}}=A_{{\tilde w}}$ and 

$A_{{\tilde y}}=A_{{w}}$.

For every solution of the infinite number of eqs-(5) by succesive iterations,  one has the $corresponding$ 

solution for the YM potentials given by eqs-(8) that are $associated$ with the 

Moyal deformations of the Killing symmetry $reductions$ of Plebanski first 

heavenly equation. Therefore,  YM potentials obtained from (5) and (8) $encode$ the 

Killing symmetry reduction. In eq-(14) we shall see that the operator equations of motion   corresponding to the Moyal quantization process of the Toda theory involves solely the operator ${\hat \Omega}$.  However, matters are not that simple because to solve the infinite number of equations (5) iteratively is far from trivial.  

The important fact is that in principle one has a systematic way of 

$solving$ (2).

The authors [9] constructed solutions to the Moyal deformations of the 

$SU(2)/SL(2)$ Nahm's equations employing the Weyl-Wigner-Moyal (WWM)  map which required the use of  

$known$ representations of $SU(2)/SL(2)$ Lie algebras [22]  in terms of 

 operators acting in the Hilbert space, $L^2(R^1)$. Also known in [9] were the solutions to the classical $SU(2)/SL(2)$ Nahm equations in terms of elliptic 

functions. The ``classical'' $\hbar \rightarrow 0$ limit of the WWM quantization of the $SU(2)$ Nahm equations was equivalent to the $N\rightarrow \infty$ limit of the 

$classical$ $SU(N)$ Nahm equations and, in this fashion, hyper Kahler metrics 

of the type discussed by [13,14] were obtained.

Another important conclusion that can be inferred from [5,9] is that one can embed the WWM-quantized $SU(2)$ 

solutions of the Moyal-deformed $SU(2)$ Nahm equations found in [9] into the $SU(\infty)$ Moyal-deformed Nahm equations and have, in this way, exact quantum solutions to the Moyal deformations of the $2D$ continuous Toda molecule which was essential in the construction of the quantum self dual membrane [1]. Since a dimensional reduction of the $W_\infty \oplus {\hat W}_\infty$ algebra is the symmetry algebra of the $2D$ effective theory, algebra that was coined $U_\infty$ in [1], one can generate other quantum solutions by $U_\infty$ co-adjoint orbit actions of the special solution found by [9]. One has then recovered the Killing symmetry reductions of the Quantum $4D$ Self Dual Gravity via the $W_\infty$ co-adjoint 

orbit method [7,8].

The case displayed here is the $converse$. We do not have ( as far as we know) $SU(\infty)$ representations in $L^2(R^1)$. However, we can in principle $solve$ (5) iteratively. The goal is now to retrieve the operator coresponding to $\Omega (w,{\tilde w},q,p;\hbar)$.

The WWM formalism [17]  establishes the one-to-one map 

that takes self-adjoint operator-valued quantities, 

${\hat \Omega}(w,{\tilde w})$, living  on the $2D$ space parametrized by  coordinates, $w,{\tilde w}$,  and acting in  the Hilbert space of $L^2 (R^1)$,   

to the space of smooth functions on the phase space manifold

${\cal M}(q,p)$ associated withe real line, $R^1$. 

The map is defined :

$$ \Omega (w,{\tilde w},q,p;\hbar)

\equiv \int^\infty_{-\infty}d\xi <q-{\xi\over 2}|{\hat \Omega } (w,{\tilde w})|q+{\xi \over 2}>exp[{i\xi p\over \hbar}]. \eqno (10a)$$

Since the l.h.s of (10a) is completely determined in terms of solutions to 

eq-(2) after the iteration process in (5) and the use of the relation (6), the r.h.s is also known : the inverse transform yields the expectation values of the operator :

$$  <q-{\xi\over 2}|{\hat \Omega } (w,{\tilde w})|q+{\xi \over 2}>=

\int^\infty_{-\infty}dp ~\Omega (w,{\tilde w},q,p;\hbar)

exp[-{i\xi p\over \hbar}]. 

\eqno (10b)$$

i.e. $all$ the matrix elements of the operator ${\hat \Omega}(w,{\tilde w})$ are determined from (10b), therefore the operator ${\hat \Omega}$ can be retrieved completely. The latter operator obeys the operator analog of the zero curvature condition, eq-(14), below. The authors in [23] have discussed ways to retrieve  distribution functions, in the quantum statistical treatment of photons, as expectation values of 

a density  operator in a diagonal basis of coherent states. Eq-(10b) suffices to obtain the full operator without the need to recur to the coherent ( overcomplete) basis of states.

It is well known by now that the SDYM equations can be obtained as a zero curvature condition [24].  In particular, eq-(2).  The operator valued extension of the zero-curvature condition reads :

$$\partial_{{\tilde z}} {\hat {\cal A}}_{{\tilde y}}-

\partial_{{\tilde y}}{\hat {\cal A}}_{{\tilde z}} +

{1\over i\hbar} [{\hat {\cal A}}_{{\tilde y}},

{\hat {\cal A}}_{{\tilde z}}]=0. \eqno (11)$$

which is the WWM transform of the original Moyal deformations of the zero curvature condition :

$$\partial_{{\tilde z}}A_{{\tilde y}}({\tilde z},q,p,w,{\tilde w};\hbar)-

\partial_{{\tilde y}}A_{{\tilde z}}({\tilde y},q,p,w,{\tilde w};\hbar) 

+\{ A _{{\tilde y}}

, A _{{\tilde z}}\}_{q,p}=0. \eqno (12)$$

This is possible due to the fact that the WWM formalism, the map ${\cal W}^{-1}$ 

preserves the Lie algebra commutation relation  :

$${\cal W}^{-1} ({1\over i\hbar}[{\hat {\cal O}}^i,{\hat {\cal O}}^j]) \equiv 

\{ {\cal O}^i,{\cal O}^j \}_{Moyal}. \eqno (13)$$

The latter equations (11,12)  can be recast $entirely$ in terms of $\Omega (w,{\tilde w},q,p,\hbar)$ and the operator ${\hat  \Omega} (w,{\tilde w})$ after one recurs to the relations $A_{{\tilde z}}=A_{{\tilde w}}; A_{{\tilde y}}=A_{w}$

(9) and the dimensional reduction conditions (7) :

$\partial_{{\tilde z}}=\partial_{{\tilde w}};

\partial_{{\tilde y}}=\partial_{ w}$. 

Hence, one arrives at the $main$ result of this section   :

$${1\over i\hbar}[{\hat \Omega}_w,{\hat  \Omega}_{{\tilde w}}]={\hat 1} \leftrightarrow 

\{ \Omega_w,{\Omega}_{{\tilde w}}\}_{Moyal}=1 . \eqno (14)$$

i.e. the operator ${\hat \Omega}$ obeys the operator equations of motion encoding the quantum dynamics. The carets denote operators. The operator form of eq-(14) was possible due to the fact that the first two terms in the zero curvature condition (12) are 

:

$$\partial_{{\tilde z}}A_{{\tilde y}}

-\partial_{{\tilde y}}A_{{\tilde z}}=-1. \eqno (15)$$

as one can verify by inspection from the dimensional reduction conditions in 

(7) and after using (8).

The operator valued expression in (14) encodes the Moyal quantization of the 

continuous Toda field.  

The original continuous Toda equation  is [2,3,18]:

$${\partial^2 \rho \over \partial {z_+} \partial z_-} ={\partial^2 e^\rho \over \partial t^2}.

~~~\rho =\rho (z_+,{z_-},t). \eqno (16) $$ 

At this stage we should point out that one should $not$ confuse the variables $z_+,{z_-},t $ of eq-(16) with the previous $z,{\tilde z}$ coordinates and the ones to be discussed below. The operator form of the Moyal deformations of (16) 

may be obtained from the ( nontrivial) Legendre map 

which takes $\Omega (w,{\tilde w},q,p;\hbar )$ to 

the Moyal-deformed $2D$ continuous Toda field. $\rho(t,\tau ,q',p';\hbar )$ 

defined in the next section. 

A further dimensional reduction  of the $3D$ continuous Moyal-Toda equation corresponds to  the deformed $2D$ continuous Toda molecule 

equation which can be embedded into the Moyal deformations of the  $SU(\infty)$ Nahm's equations.

The supersymmetric extensions follow from the results of [4] where we wrote down the Plebanski analog of $4D$ Self Dual  Supergravity .

To conclude this section : A WWM formalism is very appropriate to Moyal quantize the continuous Toda theory which we believe is the underlying theory behind 

( a sector of)  the self dual membrane. 

Due to the variable entanglement of the original Toda equation, given by the 

first equation  in  the series of eqs-(5), one has to use the dictionary relation (6) that allows to use the WWM formalism of [9] in a straightforward fashion.

In the next section we will write down reductions of the $SU(\infty)$ Moyal Nahm

equations that are linked to the continous Moyal Toda equations and discuss how

the master Legendre transform can be constructed from $\Omega$ to $\rho$.

\bigskip

\centerline{\bf CONCLUDING REMARKS}

After having Moyal-quantized the continuous Toda theory via the Moyal-Plebanski equations and

the $SU(\infty)$ Moyal-Nam equations, which in essence amounts to a Moyal deformation of

continuum Lie algebras [35] it is worth pointing out the crucial role that this will have in the

sought after membrane quantization.

The Toda theory also appears in the construction of noncritical $W_\infty$ strings . It was shown

[1] that the expected critical dimension of the (super) membrane $D=11,27$, was closely related to

the number of spacetime dimensions 

of an anomaly-free noncritical ( super) $W_\infty$ string . A BRST analysis revealed that a very

special sector of the membrane spectrum should have a relationship to the first unitary minimal

model of a ( super) $W_N$ algebra adjoined to a critical ( super) $W_N$ string spectrum in the

$N\rightarrow \infty$ limit. The study of the bosonic and supersymmetric case  furnished $D=27,11$

respectively. Noncritical $W_N$ strings are constructed by coupling $W_N$ conformal matter to $W_N$

gravity. By integrating out the matter sector the effective induced $W_N$ gravity action, in the

conformal gauge,  is obtained and it takes precisely the same form of a Toda action for the scalar

fields [26]. The same action can be obtained from a constrained WZNW model by a quantum

Drinfeld-Sokolov reduction process of a $SL(N,R)$ Kac-Moody algebra at level $k$. Each of this Toda

actions posseses a $W_N$ symmetry.

It is important to emphasize that the effective ( super) Toda theory associated with noncritical (

super) $W_\infty$ strings in $D=27,11$ dimensions  is not 

 the same Toda theory which appears in the embedding process of the continuous Toda theory into the

 $D=4$ $SU(\infty)$ Nahm equations [1]. However, one may  fix, assign or choose the coupling

 constant of the latter Toda theory in terms of the former Toda theory if one wishes [1] with the

 proviso that the $4D$ theory is anomaly free as well ( this needs to be verified).  In this case

 the coupling constant turns out to be imaginary. Imaginary couplings are typical of Affine Toda

 theories [27] that correspond to massive but integrable deformations of conformal two-dim theories.

 Soliton solutions have been found and evidence of the 

conjectured relation between solitons in one Toda theory and fundamental particles of the $dual$

Toda theory, at the quantum level, has been verified as well.

For real couplings, another duality has also been found. The $S$ matrix of the Toda theory for one

Kac-Moody algebra at coupling $\beta$ is equal to the $S$ matrix of the $dual$ Toda theory for the

dual Kac-Moody algebra with coupling 

$(4\pi/\beta)$. 

For these reasons, the Toda models may be an appropriate laboratory to test  the conjecture

dualities of string, M,F,..theory.

The masses of the solitons of these Affine Toda models can be understood in terms of a Higgs-like

mechanism associated to the spontaneous breakdown of the conformal symmetry of a more general

theory, the so-called Conformal Affine Toda models (CAT) [28] . Massless solitons travelling with

speed $less$ than $c$ have been found [29]. These CAT models can also be obtained by Hamiltonian

reduction of the two-loop WZNW models [30]. Kac-Moody extensions of the area-preserving diffs

algebra of the membrane have been constructed [31] and a Chern-Simmons gauge theory with gauge

fields valued in an affine Kac-Moody algebra, after a Hamiltonian reduction, yields also the CAT

models [32]. Once again, the 

connection between membrane and Toda models can be established via these two-loop WZNW models.

We remarked  in [1] that a Killing symmetry reduction of the  

$4D$ Quantized Self Dual Gravity ,  via the $W_\infty$ co-adjoint orbit method performed by [7,8] ,

gives a quantized Toda theory. In this letter we have presented a more direct quantization method and

quantize the Toda theory using the Weyl-Wigner-Moyal prescription (WWM). 

As emphasized in [9], BPS monopoles are solutions of the Bogomolny equations whose role has been

very relevant in the study of $D~3$-branes realizations of $N=2~D=4$ super YM theories in IIB

superstrings [10]; $D$ instantons constructions [11]; in the study of moduli spaces of BPS monopoles

and origins of ``mirror''  symmetry 

in $3D$ [12]; in constructions of self dual metrics associated with hyper Kahler spaces [13,14],

among others. A lot remains ahead. Hopefully deformation quantization techniques will be

powerful tools to tackle these very difficult problems.

\centerline {\bf Acknowledgements}

We are indebted to I. Strachan , G. Chapline and J. Plebanski for helpful discussions.

\smallskip

\centerline {\bf REFERENCES}

1-C. Castro : Phys. Lett {\bf B 288} (1992) 291.

Journ. Chaos, Solitons, Fractals, {\bf 7} no.7 (1996) 711.

``The Noncritical $W_\infty$ String Sector  of the Membrane `` hep-th/9612160.

``On the exact Integrability aspects of the self dual membrane ``hep-th/9612241.

2- E.G. Floratos, G.K. Leontaris : Phys. Lett. {\bf B 223} (1989) 153.

3- C.D. Boyer, J.Finley III : Jour. Math. Phys. {\bf 23} (6) (1982) 1126.

Q.H. Park : Int. Jour. Mod. Physics {\bf A 7} (1991) 1415.

4- C. Castro : J. Math. Phys. {\bf 34} (1993) 681.

Jour. Math. Phys {\bf 35} no.6 (1994) 3013.

5-J. Plebanski, M. Przanowski : Phys. Lett {\bf A 219} (1996) 249.

6- The original proof was given by :J.Hoppe :``Quantum Theory of a Relativistic

 Surface `` M.I.T Ph.D thesis  (1982).

E. Bergshoeff, E. Sezgin, Y. Tanni, P.K. Townsend : Ann. Phys. {\bf 199} (1990) 

340.

B. de Wit, J. Hoppe, H. Nicolai : Nucl. Phys. {\bf B 305} (1988) 545.

B.Biran, E.G. Floratos, G.K. Saviddy : Phys. Lett {\bf B 198} (1987) 32.

7- G. Chapline, K. Yamagishi : Class. Quan. Grav. {\bf 8} (1991) 427.

 G. Chapline, K. Yamagishi : Phys. Rev. Lett {bf 66} (23) (1991) 3064.

G. Chapline, B. Grossman : Phys. Lett {bf B 223}  (1989) 336.

G. Chapline : Mod. Phys. Lett {\bf A 7} (22) (1992) 1959.

8-E. Nissimov, S. Pacheva : Theor. Math. Phys. {\bf 93} (1992) 274.

9- H. Garcia-Compean, J. Plebanski : `` On the Weyl-Wigner-Moyal description

of $SU(\infty)$ Nahm equations'' hep-th/9612221.

10-D. Diaconescu `` $D$ branes, Monopoles and Nahm equations `` hep-th/9608163.

11-J.S. Park : `` Monads and $D$ Instantons `` hep-th/9612096

12- A. Hanany, E.Witten : `` $IIB$ Superstrings , BPS Monopoles and

$3D$ Gauge Dynamics `` hep-th/9611230.

13-Y. Hashimoto, Y. Yasui, S.Miyagi ,T.Otsuka :``Applications of Ashtekar

gravity to $4D$ hyper-Kahler Geometry and YM instantons.``hep-th/9610069.

14-A.Ashtekar, T. Jacobson, L. Smolin : Comm. Math. Phys. {\bf 115} (1988) 631.

15-C. Castro : Phys. Lett. {\bf B 353} (1995) 201.

16-I.A.B  Strachan : Phys. Lett {\bf B 282} (1992) 63.  Jour. Geom. Phys. {\bf 21} (1997) 255.

Jour. Phys. A {\bf 29} (1996) 6117.  Private communication.

17-H. Weyl : Z. Phys. {\bf 46} (1927) 1. E. Wigner : Phys. Rev. {\bf 40} (1932) 749.

J. Moyal : Proc. Cam. Phil. Soc. {\bf 45} (1945) 99.

18-M. Saveliev : Theor. Math. Phys. {\bf 92} (1992) 457.

M. Saveliev, A. Vershik : Phys. Lett {\bf A 143} (1990) 121.

19- D.B. Fairlie, J. Nuyts : Comm. Math. Phys. {\bf 134} (1990) 413.

20- I. Bakas, B. Khesin, E. Kiritsis : Comm. Math. Phys. {\bf 151} (1993) 233.

21-C. Pope, L. Romans, X. Shen : Phys. Lett. {\bf B 236} (1990) 173.

22- K.B. Wolf : `` Integral Transform Representations of $SL(2,R)$  in Group

Theoretical Methods in Physics `` K.B. Wolf ( Springer Verlag, 1980)

pp. 526-531

23-C.L.  Metha, E.C.G. Sudarshan : Phys. Rev. {\bf 138 B} (1963) 274.

24- M.J. Ablowitz, P.A. Clarkson : ``Solitons, Nonlinear Evolution Equations

and Inverse Scattering `` London Math. Soc. Lecture Notes vol {\bf 149}

Cambridge Univ. Press, 1991. R.S Ward : Phys. Lett {\bf A 61} (1977) 81.

Class. Quan. Grav. {\bf 7} (1990) L 217.

25-T. A. Ivanova, A. D. Popov : Jour. Math. Phys {\ bf 34} (2) (1993) 674.

T. A. Ivanova, A. D. Popov : Theor. Math. Phys {\ bf 94} (2) (1993) 316.

26- J. de Boer : `` Higher Spin Extensions of Two-Dimensional Gravity ``.

Doctoral Thesis, Univ. Utrecht, Holland, 1993.

J. de Boer, J. Goeree : Nuc. Phys. {\ bf B 381} (1992) 329.

27- P. R. Johnson : `` Exact quantum S-matrices for solitons in simply laced

affine Toda field theories `` hep-th/9611117.

D. Olive, N. Turok, J. Underwood : Nucl. Phys {\bf B 409}  (1993) 509

T. Hollowood :  Nucl. Phys {\bf B 384   }  (1992) 523

H. Braden, E. Corrigan, P. Dorey, R. Sasaki : Nucl. Phys {\bf B 338} (1990) 689

G. Delius, M. Grisaru : Nucl. Phys {\bf B 441   }  (1995) 259

N. MacKay, G. Watts : Nucl. Phys {\bf B 441   }  (1995) 277

P. Dorey :  Nucl. Phys {\bf B 358 }  (1991) 654. Nucl. Phys {\bf B 374} (1992) 741.

28- O. Babelon, L. Bonora : Phys. Lett {\bf B 244} (1990) 220.

H. Aratyn, C. Constantinidis,  L. Ferreira, J. Gomes, A. Zimmerman :

Phys. Lett {\bf B 281 } (1992) 245

 H. Aratyn, C. Constantinidis,  L. Ferreira, J. Gomes, A. Zimmerman :

 Nucl. Phys.  {\bf B 406 } (1993) 727.

29- L. Ferreira, J. Miramontes, J. Sanchez-Guillen : Nucl. Phys.  {\bf B 449} (1995) 631.

30-H. Aratyn, L. Ferreira, J. Gomes, A. Zimmerman : Phys. Lett {\bf B 254} (1991) 372.

J. Maillet : Phys. Lett {\bf B 167} (1986) 401.

A. Morozov : Sov. Jour. Nucl. Phys {\bf 52} (1990) 755

31-L. Frappat, E. Ragoucy, P. Sorba, F. Thuiller : Nucl. Phys.  {\bf B 334} (1990) 250.

32- L. Bonora, M. Martellini, Y. Zhang : Phys. Lett {\bf B 253} (1991) 373.

33- C. Castro : ``$SU(\infty)$ q-Moyal-Nahm equations and quantum deformations

of the self dual membrane `` hep-th/9704031.

34- B. Fedosov : Jour. Diff. Geometry {\bf 40} (1994) 213.

M. Reuter: ``Non-Commutative Geometry of Quantum Phase Space`` hep-th/9510011

H. Garcia-Compean, J. Plebanski, M. Przanowski : `` Geometry associated with

Self-Dual Yang-Mills and the Chiral Model approaches to Self Dual Gravity ``.

hep-th/9702046.

35- C. Castro, J. Plebanski : in preparation.

\end

\centerline {Acknowledgements}

\smallskip

We thank  I.A.B Strachan for sending us the proof of how the  Moyal-Nahm equations admit reductions to the continuum Toda chain.

\smallskip

25-O.F Dayi : ``q-deformed Star Products and Moyal Brackets `` q-alg/9609023.

I.M Gelfand, D.B Fairlie : Comm. Math. Phys {\bf 136} (1991) 487.

26-G. Rideau, P. Winternitz : Jour. Math. Phys. {\bf 34} (12) (1993) 3062.

27-J.R. Schmidt : Jour. Math. Phys. {\bf 37} (6) (1996) 3062.

28-V.G. Drinfeld : Sov. Math. Doke {\bf 32} (1985) 254.

29-M. Jimbo : Lett. Math. Phys. {\bf 10} (1985) 63.

30- G.W. Delius, A. Huffmann : `` On Quantum Lie Algebras and Quantum Root

Systems ``q-alg/9506017.

G.W. Delius, A. Huffmann, M.D. Gould, Y.Z. Zhang  : `` Quantum Lie Algebras

associated with $U_q(gl_n)$ and $U_q (sl_n)$'' q-alg/9508013.

V. Lyubashenko, A. Sudbery : `` Quantum Lie Algebras of the type $A_n$ ``

q-alg/9510004.

31-M.Reuter : ``Non Commutative Geometry on Quantum Phase Space ``

hep-th/9510011.

32-E.H. El Kinani, M. Zakkari : Phys. Lett. {\bf B 357} (1995) 105.

J. Shiraishi, H. Kubo, H. Awata, S. Odake : `` A quantum deformation of the

Virasoro algebra and Mc Donald symmetric functions `` q-alg/9507034

33- C. Zha : Jour. Math. Phys. {\bf 35} (1) (1994) 517.

E. Frenkel, N. Reshetikhin : `` Quantum Affine Algebras and Deformations of the 

Virasoro and $W$ algebras `` : q-alg/9505025.

34- J. Mas, M. Seco : Jour. Math. Phys. {\bf 37} (12) (1996) 6510.

35- L.C. Biedenharn, M.A. Lohe : `` Quantum Group Symmetry and $q$-Tensor

Algebras `` World Scientific, Singapore, 1995. Chapter 3.

36- R. Floreanini, J. le Tourneaux, L. Vincent :  Jour. Math. Phys. {\bf 37}

(8) (1996) 4135.

37-G.Dito, M. Flato, D. Sternheimer, L. Takhtajan : ``Deformation quantization

of Nambu Mechanics `` hep-th/9602016.

38- S. Albeverio,  S.M. Fei : `` Current algebraic structures over manifolds ,

Poisson algebras, $q$-deformation Quantization `` hep-th/9603114.

39- A. Connes : ``Non commutative Differential Geometry `` Publ. Math. IHES

62 (1985) 41.

40- B.V. Fedosov : J. Diff. Geometry {\bf 40} (1994) 213.

H. Garcia Compean, J. Plebanski, M. Przanowski : ``Geometry associated with

self-dual Yang-Mills and the chiral model approaches to self dual gravity ``

hep-th/ 9702046

41- A. Jevicki : `` Matrix models, Open strings and Quantization of

Membranes'' hep-th/9607187.

T. Banks, W. Fischler, S.H. Shenker, L. Susskind `` M theory as

a Matrix Model, a conjecture ``hep-th/9610043.

42- A. Marshakov : `` Nonperturbative Quantum Theories and Integrable Equations

`` ITEP-TH-47/96 preprint ( Lebedev Phys. Institute)

H. Itoyama, A. Morozov : `` Integrability and Seiberg-Witten Theory''

ITEP-M 7- 95 / OU-HET-232 preprint.

43-H.W. Braden, E. Corrigan, P.E. Dorey, R. Sasaki : Nuc. Phys {\bf B 338}

(1990) 689.

L. Bonora, V. Bonservizi : Nucl. Phys. {\bf B 390} (1993) 205.

P. Christie, G. Musardo : Nucl. Phys. {\bf B 330} (1990) 465.

D. Olive, N. Turok, J. Underwood :  Nucl. Phys. {\bf B 409} (1993) 509.

T. Hollowood : Nucl. Phys. {\bf B 384} (1992) 523.

44- G.W. Delius, M.T Grisaru, D. Zanon :  Nuc. Phys {\bf B 382 } (1992 ) 365 .

45-P.E Dorey :  Nuc. Phys {\bf B 358 } (1991 ) 654 .

46- E. Frenkel : `` Deformations of the Kdv Hierarchy and related Soliton

equations `` q-alg/9511003

47- H. Braden, A. Hone : `` Affine Toda solitons of the Calogero-Moser type ``

hep-th/ 9603178.

I. Krichever, A. Zabrodin : `` Spin generalization of the Ruijsenaars-Schneider 

model, nonabelian $2D$ Toda chain and representations of the Sklyanin algebra ``

hep-th/9505039.

\end